\documentclass[12pt,pre,a4paper,preprint,superbib]{revtex4}
\usepackage{amsmath}

\begin{document}

\title{Short waves and cyclotron losses in the relativistic gyrokinetic
theory}
\author{Alexei Beklemishev}
\affiliation{Budker Institute of Nuclear Physics, 630090 Novosibirsk, Russia}
\author{and Massimo Tessarotto}
\affiliation{Department of Mathematics and Informatics, University
of Trieste, 34127 Trieste, Italy}

\begin{abstract}
Radiation damping of the motion of charged particles in relativistic,
optically thin plasmas is described within the framework of the covariant
gyrokinetic theory \cite{BT2}. It involves description of the collisionless
single-particle dynamics as well as the Vlasov and Maxwell equations both
written in the covariant formulation. The damping causes corrections to the
phase-space trajectory of the particle, as well as to the form of the
kinetic equation itself, due to the failure of conditions of the Liouville
theorem. Both effects result independent of the gyrophase, which is retained
as an ignorable variable. In addition, the applicability range of the
covariant gyrokinetic theory is extended to describe short-wavelength
perturbations with the background of zero parallel electric field. The
presented theory is suitable for description of magnetized, relativistic,
collisionless plasmas in the context of astrophysical or laboratory
problems. Non-uniquenes of the gyrokinetic representation and consequences
thereof are discussed.

\end{abstract}

\maketitle

\section{Introduction}

Gyrokinetic description of hot magnetized plasmas\cite{LJ0} is an effective
tool that allows to combine relative simplicity of the drift approximation
with adequate account for the finite-Larmor-radius (FLR) effects, including
description of short-wavelength perturbations. Currently, most applications
are in the tokamak physics,\cite{LJ,BO,CO,Brizard} where it is used for
analysis of transport, stability and turbulence. Earlier relativistic
modifications of the theory\cite{LJ0,Hahm} were designed for simulation of
confinement of fusion products. Meanwhile, the covariant, four-dimensional
formalism of the gyrokinetic (GK) theory, developed in our previous papers
\cite{BT,BT1,BT2}, aims at being capable of describing fast flows of matter,
such as taking place in accretion disks and relativistic jets in the realm
of astrophysics. In particular, it is consistent with the general
relativity, permits the use of arbitrary four-dimensional coordinate systems
and non-canonical phase-space variables. Since four-dimensional formulation
utilizes internal symmetries of electrodynamics, so that the applicability
range of the gyrokinetic theory extends even in its non-relativistic limit.

Four-dimensional representation of drift trajectories, and its relationship
to symmetries of the Faraday tensor were first published by Fradkin.\cite{Fr}
On that basis, using the Lie-transform methods, Boghosian\cite{Bog}
constructed the first version of the covariant, four-dimensional gyrokinetic
transformation. Still, that version was not covariant in the
general-relativistic sense, inherited several restrictions of the
non-relativistic theories, and was too complicated to be of practical use.
Another formulation of covariant gyrokinetic theory was launched by
Beklemishev and Tessarotto\cite{BT} using a perturbative Lagrangian approach
introduced by Littlejohn.\cite{LJ1} The theory was just first-order in the
Larmor radius, but it was simple, straightforward and fully covariant. An
important point of that paper was concerned the non-uniqueness of the
gyrokinetic transformation, on which we are going to elaborate further
below. Additional work in this direction\cite{BT1,BT2} extended analysis to
second order in the Larmor radius and in the amplitude of the wave-field,
added explicit solutions for particle trajectories, determined the covariant
gyrokinetic equation, and the effective source-term for the Maxwell
equations.

The covariant GK theory, published in \cite{BT2} has two important
limitations, which we aim at amending in the present paper. The first is
typical to all existing gyrokinetic theories, namely, their inability to
account for cyclotron losses. The losses, while relatively small in the
non-relativistic limit, become significant for relativistic particles. In
both cases they destroy the most important integral of motion exploited by
GK theories, the \textquotedblleft magnetic moment\textquotedblright . The
effect, however, is difficult to account for within the context of any
Lagrangian formulation, since it destroys the Hamiltonian nature of the
system. Hence its inclusion in the context of GK theory requires a specific
treatment.

Radiation damping of the motion of a charged particle is a natural process
described in textbooks.\cite{LL1} However, there is a question whether it
can be directly applied for plasma particles as is. Indeed, the effect is
resonant and definitely falls out of the scope of the theory for cold dense
plasmas, where the cyclotron emission is trapped and becomes a collective
process. This statement can be illustrated by the following example: a
single particle moving in a circle emits waves, while a steady circular
current (or lots of particles moving in the same circle) emits nothing.
Well, the opposite limit is also realistic. In hot plasmas relativistic
effects destroy the resonance, so that the radiation losses grow and lose
coherence, i.e., occur as if each particle is independent. This is the case
we are going to explore: a hot rarefied plasma, with a small optical depth
in the cyclotron/synchrotron frequency range. The \textquotedblleft small
optical depth\textquotedblright\ clause is included so that the radiation
can escape, and the inverse processes of scattering/absorption can be
neglected.

In a hot rarefied plasma, the radiation damping causes corrections to the
phase-space trajectories of particles, as well as to the form of the kinetic
equation itself, since the phase-space volume is no longer conserved. We
describe these effects as small corrections, evaluating them in the
unperturbed gyrokinetic variables. Both effects result independent of the
gyrophase, which thus can be retained as an ignorable variable.

The second limitation of the covariant GK theory developed in \cite%
{BT,BT1,BT2}, which we intend to overcome in this paper, is its inability to
treat the short-wavelength perturbations when the background parallel
electric field is zero. This effectively limits description of typical
plasmas in fusion devices, where the parallel electric field is small, to
long-wavelength perturbations. In contrast, all alternative formulations of
the gyrokinetic theory are applicable \emph{only} if the electric field is
small. By searching for the gyrokinetic transformation in a broader class
than before, we are able to derive an alternative formulation, which is
valid in all limits, i.e., both for small and for large (of the order of the
magnetic field) parallel electric fields.

We also try to look at the problem of alternative formulations of the
gyrokinetic theory in broader context: is the gyrokinetic description
unique? Are all alternative formulations equivalent in some sense in their
common applicability range? We intend to show that non-equivalent
descriptions are possible and still valid in certain orders of the
perturbation theory. This feature is not related to the Lagrangian approach
but is common to all perturbative methods.

Direct comparison of two non-equivalent gyrokinetic theories, in the common
applicability range, shows that even such basic feature as drift
trajectories of the guiding center are affected by the choice of free
parameters. Exploration of these degrees of freedom provides new insight
into the nature and applicability of gyrokinetic theories. For example, a
closed drift trajectory in one representation will not necessarily be closed
in the other one. The same applies for drift surfaces in magnetic traps as
well. This feature is very important for theoretical interpretation of
cross-field transport processes. Indeed, it follows that though existence of
closed drift surfaces means that particles are contained, the reverse is not
true - those closed surfaces might still exist in other representations of
the drift variables.

The paper is organized as follows. In Section II we start by introducing
necessary notations and definitions, and in Section III present the
derivation of the covariant gyrokinetic transformation, valid for all values
of the electric field, and the relevant equations of motion. On this basis,
in Section IV we discuss the uniqueness and convergence or the gyrokinetic
approximation. Next, in Section V, we describe the effect of the radiation
damping on the motion of a particle in the gyrokinetic variables. In Section
VI the covariant kinetic and gyrokinetic equations are derived with due
account for the radiation damping effects. The section also contains our
proof of the Liouville theorem in Lagrangian variables, and explicit
expressions for the rate of contraction of the phase-space volume. Results
are summarized in Conclusion.

Throughout the paper the following notations are used: for four-dimensional
vectors we use bold letters, for tensor fields and operators use roman and
calligraphic fonts. Normal math letters are either scalars, or components of
vector and tensor fields.

\section{Coordinates, the tetrad and the variational principle}

In order to carry out the program indicated in the introduction we first
introduce the phase-space coordinates, and the tetrad of unit four-vectors
linked to symmetry planes of the Faraday tensor. In the covariant
formulation the tetrad plays a role similar to magnetic coordinates. In
these terms we formulate the variational principle for description of the
single particle dynamics. Finally, we introduce and discuss the template and
the ordering, which will be used later for construction of the gyrokinetic
transformation of the phase-space variables.

Covariance of the representation is defined as follows: suppose we have
arbitrary, user-defined coordinate system, of which we know the metric
tensor, $g_{\mu \nu }$, and the time-like invariant interval $\mathrm{d}s$
is defined as
\begin{equation}
\mathrm{d}s^{2}=g_{\mu \nu }\mathrm{d}x^{\mu }\mathrm{d}x^{\nu },  \label{ds}
\end{equation}%
where the Greek indices are assumed to go through $\mu ,\nu =0,...,3$. The
signature of $g_{\mu \nu }$ is assumed to be $<1+,3->$, so that the zero
coordinate is also time-like, and $s$ is the same as the \textquotedblleft
proper time\textquotedblright .The four-velocity $\mathbf{u}$ is then
defined in terms of its countervariant components according to
\begin{equation}
u^{\mu }=\frac{dx^{\mu }}{ds}.  \label{u}
\end{equation}%
As a result of (\ref{ds}), $\mathbf{u}$ is defined as the unit 4-vector
tangent to the trajectory. It is time-like,%
\begin{equation}
u_{\mu }u^{\mu }=1.  \label{uu=1}
\end{equation}

Thus, for an arbitrary coordinate system we have defined the
seven-dimensional phase-space as $(x^{\mu },u_{\nu })$. If, then, we present
an expression for a particle trajectory in this phase-space, it will be
covariant, i.e., valid in any coordinates. Section VI contains more
information about the phase space and distribution functions in it.

\subsection{The basis tetrad}

In the following we shall assume that in each point of space-time, {$\mathbf{%
x}$}, there is defined a \textquotedblleft tetrad\textquotedblright , i.e.,
an orthogonal basis of unit 4-vectors $\mathbf{(\tau ,l,l}^{\prime }\mathbf{%
,l}^{\prime \prime }\mathbf{)}$, such that the last three 4-vectors are
space-like, and
\begin{equation}
\epsilon _{\varsigma \lambda \mu \nu }\tau ^{\varsigma }l^{\lambda
}l^{\prime \mu }l^{\prime \prime \nu }=1,  \label{orderb}
\end{equation}%
where $\epsilon _{\varsigma \lambda \mu \nu }$\ is the purely antisymmetric
tensor. Orientation of the tetrad is arbitrary and is defined by 6
sufficiently smooth scalar functions (3 pure space rotations and 3
space-time rotations). The other 10 (out of 16 components of the tetrad
vectors) fully describe properties of the space-time itself. This feature is
used in the so-called \textquotedblleft tetrad formalism\textquotedblright\
of general relativity \cite{LL1}. It is important not to confuse this tetrad
with the basis vectors of the coordinate system.

A special choice of orientation links the tetrad $(\mathbf{\tau ,l,l}%
^{\prime }\mathbf{,l}^{\prime \prime })$ to the electromagnetic field
tensor, $F_{\mu \nu }$. We shall refer to such basis choice as
\textquotedblleft field-related\textquotedblright . Our subsequent use of
the tetrad makes other choices possible in the first approximation,\cite{BT}
but in higher orders just the field-related choice of orientation is
compatible with the solution,\cite{BT1} so we shall use it from the
beginning.

With the field-related tetrad the $(\mathbf{l}^{\prime }\mathbf{,l}^{\prime
\prime })$-plane coincides with the space-like invariant plane of the
antisymmetric tensor $F_{\mu \nu },$ while $(\mathbf{l,\tau })$ result
belonging to its other invariant plane, so that the Faraday tensor can be
fully expressed as \cite{BT1}
\begin{equation}
F_{\mu \nu }=H\left( l_{\nu }^{\prime }l_{\mu }^{\prime \prime }-l_{\mu
}^{\prime }l_{\nu }^{\prime \prime }\right) +E\left( l_{\mu }\tau _{\nu
}-l_{\nu }\tau _{\mu }\right) ,  \label{Fs}
\end{equation}%
where $H$ and $E$ are scalar functions, having the physical meaning of the
magnetic and electric fields in the reference frame where they are parallel.
Thus, the field-related basis orientation is a generalization of the
magnetic coordinates\ of the non-relativistic treatments. Fradkin\cite{Fr}
first used the invariant planes of the electromagnetic field tensor for
decomposition of motion of a charged particle.

The bivector combinations in the expression for $F_{\mu \nu },$ namely
\begin{equation}
b_{\mu \nu }=l_{\nu }^{\prime }l_{\mu }^{\prime \prime }-l_{\mu }^{\prime
}l_{\nu }^{\prime \prime },\;\;c_{\mu \nu }=l_{\mu }\tau _{\nu }-l_{\nu
}\tau _{\mu },  \label{bd}
\end{equation}%
\ can be viewed as linear operators, which combine projection on an
invariant plane with an orthogonal turn in it. Then the Faraday tensor is
\begin{equation}
\mathrm{F}=H\mathrm{b}+E\mathrm{c,}  \label{Ft}
\end{equation}%
while its dual is
\begin{equation*}
\mathcal{F}=-E\mathrm{b}+H\mathrm{c.}
\end{equation*}%
The pure projection operators $\mathcal{O}^{(a)},\mathcal{O}^{(b)}$\
introduced by Fradkin can be expressed as%
\begin{equation*}
\mathcal{O}^{(a)}=-\mathrm{bb},\;\mathcal{O}^{(b)}=\mathrm{cc}.
\end{equation*}

Since $\mathcal{O}^{(a)}+\mathcal{O}^{(b)}=\mathcal{I},$\ and {$\mathrm{b}$}$%
\mathrm{c}=0,$\ it is easy to check that the inverse of the Faraday tensor, $%
D^{\nu \mu }=(F_{\mu \nu })^{-1}$\ exists for $HE\neq 0,$ and looks like%
\textbf{\ }%
\begin{equation}
\mathrm{D}=-\frac{1}{H}\mathrm{b}+\frac{1}{E}\mathrm{c}.
\end{equation}

\subsection{Variational principle and the ordering scheme}

Particle dynamics in phase-space can be described on the basis of a
variational principle.\cite{CA} Of course, it is not unique, since different
functionals may have coinciding extrema. One of the simplest-looking forms
of such functional was used by Beklemishev and Tessarotto\cite{BT} for
description of particle dynamics in the seven-dimensional phase-space in the
framework of the general relativity. It allows to describe the particle
trajectory in the phase-space as a seven-dimensional extremal curve. In the
non-relativistic limit the extra, seventh dimension is just time, and the
whole representation can be easily reduced to conventional terms.

The phase-space trajectory of a particle with the rest-mass $m_{a}$ and
charge $q_{a}$ in prescribed fields can be found from the variational
principle $\delta S=0$, as the extremal of the functional
\begin{equation}
S{=}\int {Q_{\mu }\mathrm{d}x^{\mu }=}\int (qA_{\mu }(x^{\nu })+u_{\mu })%
\mathrm{d}{x^{\mu }},  \label{S1}
\end{equation}%
where $q=q_{a}/m_{a}c^{2}$, and $A_{\mu }$ is the four-vector potential of
the electromagnetic field. Ideally, a variational principle in the phase
space should yield the standard relativistic equation of motion of the
particle, plus the relationship between the velocity and displacement, (\ref%
{u}). As shown in \cite{BT}, this is indeed the case for the functional (\ref%
{S1}), but only for variations of $u_{\mu }$ occurring on the hypersurface $%
u_{\mu }u^{\mu }=1$. In this sense the phase space is only
seven-dimensional. The above variational principle is independent of the
parameterization of world lines. Furthermore, the above variational
principle, as any Lagrangian variational principle, is gauge invariant,
i.e., any change of the generating differential 1-form of the type
\begin{equation}
Q_{\mu }\mathrm{d}x^{\mu }\rightarrow Q_{\mu }\mathrm{d}x^{\mu }+\mathrm{d}F,
\label{1f}
\end{equation}%
where $F$ is an arbitrary smooth function of coordinates $x^{\mu }$ and $%
u_{\nu }$, does not change the extremal curve.

In general terms one can define the gyrokinetic transformation as such
transformation of the phase-space variables, that one of the new variables,
called the gyrophase, becomes ignorable, i.e., the functional, when
expressed in a suitable gauge, becomes independent of it. Then, the
canonically conjugate variable to the gyrophase, called the
\textquotedblleft magnetic moment\textquotedblright\ or the
\textquotedblleft adiabatic invariant\textquotedblright\ depending on the
context, is an integral of motion. As a result, the motion of particles in
the new gyrokinetic variables is simplified, since the system is effectively
integrated in one degree of freedom, eliminating the highest characteristic
frequency. Once the transformation and equations of motion are found, it is
possible to simplify the kinetic description as well, i.e., find the
gyrokinetic equation (see Sect.VI).

Construction of the gyrokinetic transformation is done by way of expansion
in powers of formal small parameters $\varepsilon $ and $\lambda $. They are
introduced according to the \textquotedblleft gyrokinetic\textquotedblright\
ordering scheme , which in our case, following the notation of \cite{LJ}, is%
\begin{equation}
{Q_{\mu }\mathrm{d}x^{\mu }}=\{u_{\mu }+q(\frac{1}{\varepsilon }A_{\mu }(%
\mathbf{x})+\lambda a_{\mu }(\mathbf{x}/\varepsilon ))\}\mathrm{d}{x^{\mu }},
\label{L2}
\end{equation}%
so that $\varepsilon $ accounts for the relative strength and inhomogeneity
of the background field, while $\lambda $ allows distinction between the
large-scale background field $A_{\mu },$ and the wave-fields given by $%
a_{\mu }$. In the final expressions both parameters should be set to $1$.

For an infinitesimal Larmor radius, $\varepsilon ,\lambda \rightarrow 0$,
the particle trajectory is a very tight spiral, which allows to represent
the transformation as%
\begin{equation}
x^{\mu }=x^{\prime \mu }+\sum_{s,p}\varepsilon ^{s}\lambda ^{p}r_{sp}^{\mu
}(y^{i}),  \label{xm1}
\end{equation}%
and%
\begin{equation}
u_{\nu }=\bar{u}_{\nu }+w\left( l_{\nu }^{\prime }\cos \phi +l_{\nu
}^{\prime \prime }\sin \phi \right) ,  \label{un1}
\end{equation}%
where%
\begin{equation}
\left( \bar{u}_{\nu },w\right) {=}\left( \bar{u}_{\nu }^{\prime },w^{\prime
}\right) +\sum\limits_{s,p}\varepsilon ^{s}\lambda ^{p}\left( \tilde{u}_{\nu
},\tilde{w}\right) _{sp}.  \label{un2}
\end{equation}%
Here $y^{i}=(x^{\prime \mu },\phi ,\mu ,u_{\Vert })$ are the new gyrokinetic
variables; $x^{\prime \mu }$ being the coordinate of the spiral center, the
gyrocenter; the Larmor corrections $r_{s}^{\mu }(y^{i})$ describe the
difference between the particle position and its gyrocenter, and are thus
defined to be purely oscillatory in the gyrophase, $\phi $; $\mathbf{l}%
^{\prime }$ and $\mathbf{l}^{\prime \prime }$ are the tetrad vectors
defining the \textquotedblleft plane of rotation\textquotedblright , they
are assumed to vary on the scale of the background field and depend on $%
\mathbf{x^{\prime }}$ only; $\bar{u}_{\nu }^{\prime }$ is the average part
of velocity, $w^{\prime }$ is its \textquotedblleft
perpendicular\textquotedblright\ component, while oscillatory corrections to
the velocity components describe the possibility that the velocity along the
Larmor orbit may behave in a complex fashion in higher orders.

The transformation (\ref{un1}),(\ref{un2}) should preserve the velocity as a
unit vector. This requirement, expressed by Eq.(\ref{uu=1}), is satisfied if
$\bar{u}_{\nu }$ is orthogonal to $\mathbf{l}^{\prime }\mathbf{,l}^{\prime
\prime }$, i.e.,%
\begin{equation}
\bar{u}_{\nu }=u_{o}\tau _{\nu }+{u_{\parallel }l}_{\nu },  \label{un3}
\end{equation}%
while parameters $w,{u_{\parallel },}u_{o}$ satisfy%
\begin{equation}
u_{o}^{2}=1+w^{2}+{u_{\parallel }}^{2},  \label{uo}
\end{equation}%
which follows from $u_{\mu }u^{\mu }=1$ for all $\phi $. Note that $u_{o}$
is \emph{not} the $0$th component of $\mathbf{u}.$

There are just 6 independent corrections to be defined in each order of the
expansion. It is possible to introduce the seventh free function by way of a
correction to $\phi ,$ but it is not necessary. Thus, the gyrophase $\phi $
is defined as\textit{\ }\emph{an angle in the velocity-subspace}\textit{, }%
where we introduced a cylindrical coordinate system linked to the Faraday
tensor. This definition is covariant with respect to transformations of the
space-time coordinate system, which may change the vector components, but
not the vectors themselves.

As compared to our previous work,\cite{BT1} the gyrokinetic transformation
is being searched for in broader class of transformations, (\ref{xm1})-(\ref%
{un2}), than before. Our previous formulation left components of the
velocity unperturbed, i.e., $\left( \tilde{u}_{\nu },\tilde{w}\right)
_{sp}\equiv 0.$ This made the whole work simpler, but there was no solution
in one particular limit in orders $\lambda ^{i},i>1.$ Without the wave-field
($\lambda ^{0}$) solution existed for all parameters.

A more natural way of defining the velocity transformation would be to keep $%
\bar{u}_{\nu }$ independent of $\phi $, i.e., discard Eq.(\ref{un2}), but
allow the tetrad to flex with respect to the background field tensor.
Unfortunately, this involves too many variables, and is not realized at the
moment.

The ordering scheme should ensure that all components of the particle
velocity are of the same order despite the ordering of displacement, (\ref%
{xm1}). This is achieved by ordering the $\mathrm{d}x^{\mu }$ vector
components in the new variables $y^{i}$ as
\begin{equation}
\mathrm{d}x^{\mu }=\frac{\partial x^{\mu }}{\partial y^{i}}\mathrm{d}{y^{i}}%
=\left( \mathrm{d}x^{\mu }\right) _{\phi }+\frac{1}{\varepsilon }\frac{%
\partial x^{\mu }}{\partial \phi }\mathrm{d}\phi .  \label{dxmuprime}
\end{equation}%
Here $\left( \mathrm{d}x^{\mu }\right) _{\phi }$ means the differential of $%
x^{\mu }$ with respect to all variables of the new set $y^{i}$ except $\phi $%
.

The ordering is designed to ensure that the expansion into the Taylor series
in $(x^{\mu }-x^{\prime \mu }),$ i.e., in the Larmor radius, is possible for
components of the background field, $A_{\mu }$, and the orientation vectors,
$l^{\prime \mu }$ and $l^{\prime \prime \mu }.$ Note, that we are going to
expand the vector components using partial derivatives, rather than vectors
using covariant derivatives. Thus, we are going to expand the scalar
products of vector fields by basis vectors. For this to be possible, the
metric coefficients, and, hence, the coordinate system and the curvature of
the space-time should be sufficiently smooth, i.e., vary on scales larger
than the Larmor radius. In contrast, the wave-field, $a_{\mu }(\mathbf{x}%
/\varepsilon ),$ is allowed to vary rapidly, so that its Taylor expansion in
$(x^{\mu }-x^{\prime \mu })$ is impossible. Instead of the Taylor expansion
the Fourier representation can be used for $a_{\mu }$. Nevertheless, $a_{\mu
}$ can be still Taylor-expanded in powers of the displacement from the
\textquotedblleft unperturbed\textquotedblright\ orbit. This displacement is
proportional to the amplitude of the wave and is required to be suitably
smaller than the wavelength.\

\section{The gyrokinetic theory}

The goal of the gyrokinetic transformation is to make $\phi $ the ignorable
variable, and determine the corresponding integral of motion. We shall first
suppose that our general template of the transformation, (\ref{xm1})-(\ref%
{dxmuprime}), satisfies the requirement, apply it to the variational
principle, and then determine the arbitrary functions of the template in
such a way that the Lagrangian is indeed independent of $\phi .$ The
procedure will proceed order by order. Most of the related algebra is
already published,\cite{BT2} so that we shall drop common expressions and
highlight differences in the process of derivation.

First we substitute expressions (\ref{xm1}),(\ref{dxmuprime}) into (\ref{L2}%
) and drop terms of order $\varepsilon ^{2}$ or smaller, assuming that $%
u_{\mu }$ is of order zero, and expanding components of the vector potential
into the Taylor series. Then, following\cite{BT} we construct the
gauge-modified functional
\begin{equation}
{\delta G}^{\prime }={Q_{\mu }\mathrm{d}x^{\mu }}-\mathrm{d}{}\left( \frac{1%
}{2}q(A_{\mu }+A_{\mu }^{\prime })(r_{1}^{\mu }+\varepsilon r_{2}^{\mu
})\right) ,  \label{forr2}
\end{equation}%
where $A_{\mu }^{\prime }=A_{\mu }(x^{\prime \nu }).$ This gauge makes the
expression more symmetric and converts all references to partial derivatives
of the vector potential into components of the Faraday tensor at $\mathbf{%
x^{\prime }}$:
\begin{equation}
F{_{\mu \nu }}=\frac{\partial A_{\nu }^{\prime }}{\partial x^{\prime {\mu }}}%
-\frac{\partial A_{\mu }^{\prime }}{\partial x^{\prime {\nu }}}.
\end{equation}%
As a result, ${\delta G}^{\prime }$ becomes a long expression with terms
proportional to $\mathrm{d}{x^{\prime \mu },}\left( \mathrm{d}{r_{1}^{\mu }}%
\right) _{\phi },$ and $\mathrm{d}\phi :$%
\begin{multline}
{\delta G}^{\prime }=\left( \frac{q}{\varepsilon }A_{\mu }^{\prime }+u_{\mu }%
{+}\lambda qa_{\mu }-qr_{1}^{\nu }F{_{\mu \nu }}-\varepsilon q\left(
r_{2}^{\nu }F{_{\mu \nu }+}\frac{1}{2}r_{1}^{\nu }r_{1}^{\varsigma }\frac{%
\partial F_{\mu \nu }}{\partial x^{\prime {\varsigma }}}\right) \right)
\mathrm{d}{x^{\prime \mu }}+\hspace*{5cm} \\
+\varepsilon \left( u_{\mu }+\lambda qa_{\mu }-\frac{1}{2}qr_{1}^{\nu }F{%
_{\mu \nu }}\right) \left( \mathrm{d}{r_{1}^{\mu }}\right) _{\phi
}+\varepsilon \left( qA_{\mu }^{\prime }\frac{\partial r_{3}^{\mu }}{%
\partial \phi }+(u_{\mu }+\lambda qa_{\mu })\frac{\partial r_{2}^{\mu }}{%
\partial \phi }\right) \mathrm{d}\phi + \\
+\varepsilon \left( -\frac{1}{2}q\left( r_{1}^{\nu }\frac{\partial
r_{2}^{\mu }}{\partial \phi }+r_{2}^{\nu }\frac{\partial r_{1}^{\mu }}{%
\partial \phi }\right) F{_{\mu \nu }+}\frac{1}{2}qr_{1}^{\varsigma
}r_{1}^{\nu }\frac{\partial r_{1}^{\mu }}{\partial \phi }\frac{\partial }{%
\partial x^{\prime {\varsigma }}}\left( \frac{1}{2}\frac{\partial A_{\mu
}^{\prime }}{\partial x^{\prime {\nu }}}-\frac{\partial A_{\nu }^{\prime }}{%
\partial x^{\prime {\mu }}}\right) \right) \mathrm{d}\phi +o(\varepsilon ).
\label{dG2}
\end{multline}%
Up to this point there are no differences in the derivation.

We proceed by eliminating from ${\delta G}^{\prime }$ all terms oscillating
in $\phi .$ This is done by solving for displacements $r_{sp}^{\mu }$ and
the velocity corrections $\left( \tilde{u}_{\nu },\tilde{w}\right) _{sp}$ in
each order in $\varepsilon $ and $\lambda $.

\subsection{Order $\protect\varepsilon ^{1}$}

Initially we drop all terms in Eq.(\ref{dG2}), which are of order $%
\varepsilon ^{1}$or higher, while retaining contributions of order $\lambda $%
. Conditions for $\phi $ to be ignorable for the gauge-modified functional ${%
\delta G}^{\prime \prime }={\delta G}^{\prime }-\varepsilon \mathrm{d}R$ ,
where $R$ is an arbitrary Gauge function look like
\begin{equation}
\tilde{u}_{\mu }+\lambda q\tilde{a}_{\mu }-qr_{1}^{\nu }F_{\mu \nu } =0;
\label{cond}
\end{equation}
\begin{equation}
\left\{ (u_{\mu }+\lambda qa_{\mu }-\frac{1}{2}qr_{1}^{\nu }F_{\mu \nu })
\frac{\partial r_{1}^{\mu }}{\partial \phi }\right\} ^{\sim } =\frac{%
\partial R}{\partial \phi }.  \label{ignore}
\end{equation}
Here $\tilde{y}$ denotes the oscillating part of $y$, namely $\tilde{y}=y-%
\bar{y}$, where $\bar{y}=\langle y\rangle _{\phi }$ is the $\phi $%
-independent part of $y$. By a proper choice of the gauge-function $R$\ one
can always satisfy equation (\ref{ignore}). This means that
\begin{equation}
\tilde{u}_{\mu }+\lambda q\tilde{a}_{\mu }-qr_{1}^{\nu }F_{\mu \nu }=0
\label{co}
\end{equation}%
yields the only essential requirements. The physical meaning of Eq.(\ref{co}%
) is the relationship between the first-order gyro-radius, $\mathbf{r}_{1},$
and the oscillating velocity, $\mathbf{\tilde{u},}$ along the Larmor orbit.
The time-like component describes changes in energy due to displacements of
charge in the electric field.

If the above requirements (\ref{cond}) and (\ref{ignore}) are satisfied and $%
\phi $ is ignorable, the hybrid variational principle in our approximation
can be expressed as $\delta S^{\prime \prime }=0$ with
\begin{equation}
S^{\prime \prime }=\int \delta G^{\prime \prime }=\int \left\{ \left( \frac{q%
}{\varepsilon }A_{\mu }^{\prime }+\left\langle u\right\rangle _{\mu
}+\lambda q\overline{a}_{\mu }\right) \mathrm{d}x^{\prime \mu }+\frac{1}{2}%
\left\langle (\tilde{u}_{\mu }+\lambda q\tilde{a}_{\mu })\frac{\partial
r_{1}^{\mu }}{\partial \phi }\right\rangle _{\phi }\mathrm{d}\phi \right\}
\emph{.}
\end{equation}

Equation (\ref{co}) can be formally solved for $r_{1}^{\nu }$ to yield
\begin{equation}
r_{1}^{\nu }=\frac{1}{q}D^{\nu \mu }\tilde{u}_{\mu }+\lambda D^{\nu \mu }%
\tilde{a}_{\mu },  \label{r1m}
\end{equation}%
where $D^{\nu \mu }=(F_{\mu \nu })^{-1}$, provided that $F_{\mu \nu }$ is
not degenerate, i.e., $E\neq 0.$

Unfortunately, the degenerate case of zero parallel electric field is the
only case that is studied by all other gyrokinetic theories.\cite{LJ} Thus
we need to present a solution that would be applicable in all cases. To do
this we split equation (\ref{co}) in two projections using operators $%
\mathrm{b,c}$ . Taking its products with $b^{\lambda \mu },c^{\lambda \mu }$
we get
\begin{eqnarray}
b^{\lambda \mu }\tilde{u}_{\mu }+\lambda qb^{\lambda \mu }\tilde{a}_{\mu
}-qHr_{1}^{\lambda (b)} &=&0,  \label{projb} \\
c^{\lambda \mu }\tilde{u}_{\mu }+\lambda qc^{\lambda \mu }\tilde{a}_{\mu
}+qEr_{1}^{\lambda (c)} &=&0,  \label{projc}
\end{eqnarray}%
where $r_{1}^{\lambda (b)}$ has only $(\mathbf{l}^{\prime }\mathbf{,l}%
^{\prime \prime })$-components, and $r_{1}^{\lambda (c)}$ - only $(\mathbf{%
l,\tau }).$

We obviously do not need all components of $\tilde{u}_{\mu }$ and $%
r_{1}^{\lambda }$ to satisfy these two equations. Solution is not unique.
Our previous choice was to set $\tilde{u}_{\mu }^{(c)}$ to zero, which
becomes incompatible with equations for $E=0.$ Another possible choice,
which is applicable for all regimes, is to set $r_{1}^{\lambda (c)}=0.$
Note, that this choice is one among many with the same applicability, its
main advantage being just relative simplicity. Different choices lead to
non-equivalent gyrokinetic transformations (see discussion in Section IV,)
in particular, our current choice will make the result different from the
previous formulation\cite{BT2} even in the common applicability range.

From the $\mathrm{b}$-projection, (\ref{projb}), we immediately get

\begin{equation}
r_{1}^{\lambda }=\frac{1}{qH}\left[ w(l^{\prime \prime \lambda }\cos \phi
-l^{\prime \lambda }\sin \phi )\right] ^{\sim }+\frac{\lambda }{H}b^{\lambda
\mu }\tilde{a}_{\mu }.  \label{r1mm}
\end{equation}%
(here the superscript of $\mathbf{r}_{1}$ has been dropped since the other
projection, $\mathbf{r}_{1}^{(c)},$ is zero.) The $\mathrm{c}$-projection, (%
\ref{projc}), yields
\begin{equation}
\left( \tilde{u}_{\mu }+\lambda q\tilde{a}_{\mu }\right) \tau ^{\mu }=\left(
\tilde{u}_{\mu }+\lambda q\tilde{a}_{\mu }\right) l^{\mu }=0,  \label{atl}
\end{equation}%
and it follows that%
\begin{eqnarray*}
\tilde{u}_{o} &=&-\lambda q\tilde{a}_{\mu }\tau ^{\mu }, \\
\tilde{u}_{\Vert } &=&\lambda q\tilde{a}_{\mu }l^{\mu }.
\end{eqnarray*}%
This means that in orders $\lambda ^{0}$ (without the wave) the oscillating
parts of $w,{u_{\parallel },}u_{o}$ can be ignored. If $\lambda \neq 0,$
restriction (\ref{uo}) causes $w$ to have the oscillating part as well.

After calculating averages with the help of the above solution, the $\phi $%
-independent functional $S^{\prime \prime }$ becomes
\begin{equation}
S^{\prime \prime }=\int \left\{ \left( \frac{q}{\varepsilon }A_{\mu
}^{\prime }+\lambda q\overline{a}_{\mu }+\left\langle u\right\rangle _{\mu
}\right) \mathrm{d}x^{\prime \mu }+\widehat{\mu }\mathrm{d}\phi \right\}
\emph{,}  \label{AS}
\end{equation}%
where
\begin{equation}
\left\langle u\right\rangle _{\mu }=\bar{u}{_{\parallel }}l_{\mu }+\bar{u}%
_{o}\tau _{\mu }+\left\langle \tilde{w}\left( l_{\mu }^{\prime }\cos \phi
+l_{\mu }^{\prime \prime }\sin \phi \right) \right\rangle ,  \label{uma}
\end{equation}%
and $\hat{\mu}$ is the wave-field-modified magnetic moment, accurate to
order $\varepsilon ^{1}$,
\begin{equation}
\hat{\mu}=\frac{\left\langle w^{2}\right\rangle }{2qH}+\frac{\lambda }{H}%
\left\langle \tilde{a}_{\nu }b^{\nu \mu }\frac{\partial }{\partial \phi }%
\left[ w(l_{\mu }^{\prime }\cos \phi +l_{\mu }^{\prime \prime }\sin \phi )+%
\frac{1}{2}\lambda q\tilde{a}_{\mu }\right] \right\rangle .  \label{MMF}
\end{equation}%
Note that only $\mathrm{b}$-projections of the wave-potential enter this
definition, unlike the previous formulation, where all components were
involved.

The evaluation of the gyrophase averages involving the wave-field $a_{\mu }$
is, of course, the difficult part. The reason for this is the necessity to
transform the given function of space-time coordinates $a_{\mu }(x^{\nu })$
into a function of new variables, while the transformation rule (\ref{r1m}),(%
\ref{r1mm}) is itself dependent on $a_{\mu }$. Solution of the equation
\begin{equation}
a_{\mu }=a_{\mu }\left( x^{\prime \lambda }+\frac{\varepsilon }{qH}\left[
w(l^{\prime \prime \lambda }\cos \phi -l^{\prime \lambda }\sin \phi )\right]
^{\sim }+\frac{\lambda }{H}b^{\lambda \mu }\tilde{a}_{\mu }\right)
\label{atransform}
\end{equation}%
can be only obtained by expanding in powers of $\lambda $.

\subsection{Order $\protect\varepsilon ^{1}\protect\lambda ^{0}$}

Solution of the problem in the limit $\lambda =0$ is simple \cite{BT}: since
the oscillating parts of $w,{u_{\parallel },}u_{o}$ can be ignored,
\begin{equation}
\hat{\mu}\equiv \frac{w^{2}}{2qH},  \label{mudef}
\end{equation}%
can be interpreted as the magnetic moment. By varying the functional (\ref%
{AS}) with respect to $\phi $, we find $\mathrm{d}\hat{\mu}=0,$ i.e., $\hat{%
\mu}$ is the first integral of motion in this order.

Now the functional (\ref{AS}) can be rewritten as
\begin{equation}
S^{\prime \prime }=\int \left\{ \left( \frac{q}{\varepsilon }A_{\mu
}^{\prime }+u{_{\parallel }}l_{\mu }+u_{o}\tau _{\mu }\right) \mathrm{d}%
x^{\prime \mu }+\hat{\mu}\mathrm{d}\phi \right\} ,  \label{F1}
\end{equation}%
with $u_{o}=\sqrt{1+2qH\hat{\mu}+{u_{\parallel }}^{2}}$.

\subsection{Order $\protect\lambda ^{0}\protect\varepsilon ^{2}$}

This order of the perturbation theory is less trivial, but, due to vanishing
$\tilde{w},{\tilde{u}_{\parallel },}\tilde{u}_{o}$, coincides with the
derivation provided in our previous paper, \cite{BT2}. An important feature
arising in this order is the solubility condition,
\begin{equation}
\left\{ \left( {u_{\parallel }}l_{\mu }+u_{o}\tau _{\mu }\right) D^{\mu \nu
}(l_{\nu }^{\prime }\cos \phi +l_{\nu }^{\prime \prime }\sin \phi )\right\}
^{\sim }=0,
\end{equation}%
which is satisfied identically if the $(\mathbf{l}^{\prime }\mathbf{,l}%
^{\prime \prime })$-plane is the invariant plane of $F^{\mu \nu }$. This is
the only requirement that motivates the \textquotedblleft
field-related\textquotedblright\ choice of the basis $(\mathbf{\tau ,l,l}%
^{\prime }\mathbf{,l}^{\prime \prime }).$

In this order the $\phi $-independent functional takes the form%
\begin{equation}
S^{\prime \prime }=\int \left\{ \left( \frac{q}{\varepsilon }A_{\mu
}^{\prime }+{u_{\parallel }}l_{\mu }+u_{o}\tau _{\mu }+\frac{1}{2}%
\varepsilon \hat{\mu}\chi _{\mu }\right) \mathrm{d}{x^{\prime \mu }}+\hat{\mu%
}\mathrm{d}\phi \right\} ,
\end{equation}%
where the previous expression for the adiabatic invariant, (\ref{mudef}),
holds. Thus, the second order modifications to the Lagrangian can be
described as inhomogeneity contributions to the effective electromagnetic
potential with
\begin{equation}
\chi _{\mu }=l_{\nu }^{\prime }\frac{\partial l^{\prime \prime \nu }}{%
\partial {x^{\prime \mu }}}-l_{\nu }^{\prime \prime }\frac{\partial
l^{\prime \nu }}{\partial {x^{\prime \mu }}}-(l^{\prime \nu }l^{\prime
\varsigma }+l^{\prime \prime \nu }l^{\prime \prime \varsigma })\frac{1}{H}%
\frac{\partial F_{\mu \nu }}{\partial x^{\prime {\varsigma }}},
\label{chimu}
\end{equation}%
\ while the form of the adiabatic invariant remains unchanged. Note that
some of the effects of the gravitational field are also contained in $\chi
_{\mu }$ via derivatives of the basis vectors.

\subsection{Order $\protect\varepsilon ^{1}\protect\lambda ^{1}$}

We proceed by considering the linear approximation in the wave-field
amplitude without high-order curvature effects. As noted above, the main
problem here is the transformation of the highly-local wave-field $a_{\mu }(
\mathbf{x})$, where $\mathbf{x}$ is the particle position, to the
guiding-center coordinates $(x^{\prime \alpha },\phi ,\widehat{\mu},{%
u_{\parallel }})$. The Taylor-expansion procedure used above for
transformation of the equilibrium field fails if the wavelength is
sufficiently short. Instead, Eq.(\ref{atransform}) is solved using Fourier
analysis and expansion in powers of $\lambda $.

Assume that the wave field is given in terms of its Fourier components $%
a_{\mu }(\mathbf{k}),$ then Eq.(\ref{atransform}) becomes
\begin{equation}
a_{\mu }(\mathbf{x})=\frac{1}{\left( 2\pi \right) ^{2}}\int \mathrm{d}^{4}k\
a_{\mu }(\mathbf{k})e^{i\mathbf{kx}^{\prime }}\exp \left\{ i\varepsilon
k_{\nu }\left( \frac{w}{qH}(l^{\prime \prime \nu }\cos \phi -l^{\prime \nu
}\sin \phi )+\frac{\lambda }{H}b^{\nu \mu }\tilde{a}_{\mu }\right) ^{\sim
}\right\} ,  \label{aM}
\end{equation}%
where only the second exponential factor depends on $\phi $.

Here the ordering of the wave-vector $\mathbf{k}$ is such that $\varepsilon
k\sim O(1),$ as usual \cite{LJ}, but without additional restrictions on the
\textquotedblleft parallel wavelength\textquotedblright\ and frequency,
which appear in traditional approaches. Namely, we allow $\omega /\Omega
\sim O(1)$ and $k_{\Vert }\rho \sim O(1)$ in place of the usual ordering $%
\sim O(\varepsilon )$.

We can calculate the required averages in (\ref{AS}) and (\ref{MMF}) by
extracting the main exponential contribution and expanding it into the
Fourier series in $\phi $. We use the following identity\cite{Ab}
\begin{equation}
\exp \left[ i\zeta \sin \psi \right] =\sum_{n=-\infty }^{+\infty
}J_{n}\left( \zeta \right) \exp \left( in\psi \right) ,  \label{tojd}
\end{equation}%
to get
\begin{equation}
a_{\mu }(\mathbf{x})=\frac{1}{\left( 2\pi \right) ^{2}}\int \mathrm{d}^{4}k\
a_{\mu }(\mathbf{k})e^{i\mathbf{kx}^{\prime }}\sum_{n=-\infty }^{+\infty
}J_{n}\left( \xi \right) \exp \left( in(\phi _{0}-\phi )\right) \exp \left(
i\varepsilon k_{\nu }\delta r^{\nu }\right) ,  \label{a(x)}
\end{equation}%
where
\begin{equation}
\delta r^{\nu }=\left[ \frac{\tilde{w}}{qH}(l^{\prime \prime \nu }\cos \phi
-l^{\prime \nu }\sin \phi )\right] ^{\sim }+\frac{\lambda }{H}b^{\nu \mu }%
\tilde{a}_{\mu }  \label{drn}
\end{equation}%
characterizes the residual terms. We also defined $\xi $ as
\begin{equation}
\xi =k_{\bot }\rho =(\varepsilon \bar{w}/qH)\sqrt{\left( k_{\nu }l^{\prime
\prime \nu }\right) ^{2}+\left( k_{\nu }l^{\prime \nu }\right) ^{2}},
\label{ksi}
\end{equation}%
where $k_{\bot }=\sqrt{\left( k_{\nu }l^{\prime \nu }\right) ^{2}+\left(
k_{\nu }l^{\prime \prime \nu }\right) ^{2}}$ is the length of $\mathbf{k}%
_{\bot }$, which is the projection of $\mathbf{k}$ on the $(\mathbf{\
l^{\prime },l^{\prime \prime }})$-plane; $\phi _{0}=\arctan \left( k_{\nu
}l^{\prime \prime \nu }/k_{\nu }l^{\prime \nu }\right) $ is the angle of $%
\mathbf{k}_{\bot }$ with respect to the $(\mathbf{l^{\prime },l^{\prime
\prime }})$-basis, $\rho =\varepsilon \bar{w}/qH$ is the Larmor radius.

To the order $\varepsilon ^{1}\lambda ^{1}$,$\ \ \delta r^{\nu }$ in the
expression (\ref{a(x)}) can be set to $0$, so that its zero-order solution
is just
\begin{equation}
a_{\mu }^{(0)}(\mathbf{x})=\frac{1}{\left( 2\pi \right) ^{2}}\int \mathrm{d}%
^{4}k\ a_{\mu }(\mathbf{k})e^{i\mathbf{kx}^{\prime }}\sum_{n=-\infty
}^{+\infty }J_{n}\left( \xi \right) \exp \left( in(\phi _{0}-\phi )\right) .
\label{a0(x)}
\end{equation}

Note, that neglecting the last exponent in equation (\ref{a(x)}), or solving
by expansion in powers of $\lambda ,$ is possible only if the particle
displacement from the equilibrium orbit due to the wave field is much
smaller than the wavelength.

Now that the oscillating part of the wave-potential is defined, we can find
the oscillating parts of the velocity parameters from Eqs.(\ref{atl}),(\ref%
{uo}). In particular, in this order%
\begin{equation}
\tilde{w}^{(1)}=-\frac{\lambda q}{\bar{w}}u^{\prime \nu }\tilde{a}_{\nu },
\label{wwave}
\end{equation}%
with%
\begin{equation}
\bar{u}^{\prime \nu }=\bar{u}{_{\parallel }}l^{\nu }+\bar{u}_{o}\tau ^{\nu }.
\label{uprime}
\end{equation}

Now the averages, present in expressions (\ref{AS}), (\ref{MMF}), become
\begin{gather}
\bar{a}_{\mu }^{(0)}=\frac{1}{4\pi ^{2}}\int \mathrm{d}^{4}k\ a_{\mu }(%
\mathbf{k})e^{i\mathbf{kx}^{\prime }}J_{0}\left( \xi \right) ,  \label{am0}
\\
\left\langle (l^{\prime \prime \nu }\sin \phi +l^{\prime \nu }\cos \phi )%
\tilde{a}_{\nu }^{(0)}\right\rangle =\frac{1}{4\pi ^{2}}\int \mathrm{d}%
^{4}k\ a_{\nu }(\mathbf{k})e^{i\mathbf{kx}^{\prime }}iJ_{1}\left( \xi
\right) (l^{\prime \nu }\sin \phi _{0}-l^{\prime \prime \nu }\cos \phi _{0}),
\\
\left\langle \tilde{w}\left( l_{\mu }^{\prime }\cos \phi +l_{\mu }^{\prime
\prime }\sin \phi \right) \right\rangle =-\frac{\lambda qu^{\prime \nu }}{%
4\pi ^{2}\bar{w}}\int \mathrm{d}^{4}k\ a_{\nu }(\mathbf{k})e^{i\mathbf{kx}%
^{\prime }}iJ_{1}\left( \xi \right) (l_{\mu }^{\prime }\sin \phi _{0}-l_{\mu
}^{\prime \prime }\cos \phi _{0}).
\end{gather}%
so that to the order $\varepsilon ^{1}\lambda ^{1}$
\begin{equation}
S^{\prime \prime }=\int \left\{ \left( \frac{q}{\varepsilon }A_{\mu
}^{\prime }+\lambda q\bar{a}_{\mu }+\bar{u}_{\mu }^{\prime }+\bar{u}{_{\bot
\mu }}\right) \mathrm{d}x^{\prime \mu }+\hat{\mu}\mathrm{d}\phi \right\}
\emph{,}  \label{Sl1}
\end{equation}%
where $\bar{a}_{\mu }=\bar{a}_{\mu }^{(0)},$ and%
\begin{equation}
\bar{u}{_{\bot \mu }=}\frac{\lambda \varepsilon }{2H}b_{\mu }^{\;\lambda }%
\bar{f}_{\lambda \nu }\bar{u}^{\prime \nu },  \label{uperp}
\end{equation}

\begin{equation}
\hat{\mu}=\frac{\bar{w}^{2}}{2qH}\left( 1-\frac{\lambda \varepsilon }{H}%
b^{\mu \nu }\bar{f}_{\mu \nu }\right) ,  \label{mula}
\end{equation}%
and we have introduced the new notation%
\begin{equation}
\bar{f}_{\mu \nu }=\frac{1}{4\pi ^{2}}\int \mathrm{d}^{4}k\ i\left[ k_{\mu
}a_{\nu }-k_{\nu }a_{\mu }\right] e^{i\mathbf{kx}^{\prime }}\frac{%
J_{1}\left( \xi \right) }{\xi }=\frac{1}{4\pi ^{2}}\int \mathrm{d}^{4}k\
f_{\mu \nu }(\mathbf{k})e^{i\mathbf{kx}^{\prime }}\frac{J_{1}\left( \xi
\right) }{\xi },  \label{fbar}
\end{equation}%
thus expressing the adiabatic invariant $\hat{\mu}$ and $\bar{u}{_{\perp \mu
}}$ in terms of the gauge-invariant Faraday tensor of the wave,
\begin{equation*}
f_{\mu \nu }=\partial a_{\nu }/\partial x^{\mu }-\partial a_{\mu }/\partial
x^{\nu }.
\end{equation*}

Note that the correction in Eq.(\ref{mula}) is proportional to the averaged
parallel component of the magnetic field of the wave, $b^{\mu \nu }\bar{f}%
_{\mu \nu }$. This effect is obviously necessary, but is usually neglected
due to the standard gyrokinetic ordering of the wave-field, so that the
first order contribution to the magnetic moment vanishes.

Note also that in this order we can use the relationship%
\begin{equation}
\bar{u}_{0}^{2}-\bar{w}^{2}-\bar{u}_{\Vert }^{2}=1  \label{uo1}
\end{equation}%
instead of (\ref{uo}).

\subsection{Summary}

Finally, we summarize the above results by presenting the transformed
variational principle valid through the first order in $\lambda $\ and
second order in $\varepsilon ,$ i.e., with terms of the order $\varepsilon
\lambda $ and $\varepsilon ^{2}$ retained: $\delta S=0$ yields the particle
phase-space trajectory with
\begin{equation}
S=\int \left( qA_{\mu }^{\prime }+u{_{\parallel }}l_{\mu }+\left(
1+2qH^{\ast }\hat{\mu}+{u_{\parallel }}^{2}\right) ^{1/2}\tau _{\mu }+q\bar{a%
}_{\mu }+\bar{u}{_{\bot \mu }}+\frac{1}{2}\hat{\mu}\chi _{\mu }\right)
\mathrm{d}{\ x^{\prime \mu }}+\hat{\mu}\mathrm{d}\phi ,  \label{summary}
\end{equation}%
where $({x^{\prime \mu },}u{_{\parallel },}\hat{\mu},\phi )$ or $({x^{\prime
\mu },}u{_{\parallel },}w,\phi )$ are the new gyrokinetic variables with $%
\hat{\mu}=w^{2}/2qH^{\ast }$; $q=q_{a}/m_{a}c^{2}$ is the signed
charge-to-mass ratio;
\begin{equation}
H^{\ast }=H+b^{\mu \nu }\bar{f}_{\mu \nu },
\end{equation}%
is the effective magnetic field strength, with $\bar{f}_{\mu \nu }$ being
the averaged Faraday tensor of the wave, given by Eq.(\ref{fbar}).
Expressions for the averaged potential of the wave, $\bar{a}_{\mu },$ the
wave-induced drift, $\bar{u}{_{\bot \mu },}$ and the second-order
correction, $\chi _{\mu },$ are given by equations (\ref{am0}),(\ref{uperp}%
), and (\ref{chimu}) respectively.

The orthogonal basis $(\mathbf{\tau ,l,l^{\prime },l^{\prime \prime }})$ is
chosen in such a way that $(\mathbf{l^{\prime },l^{\prime \prime })}$
coincides with the space-like invariant plane of the antisymmetric tensor of
the electromagnetic field $F{_{\mu \nu }}=\partial A_{\nu }^{\prime
}/\partial x^{\prime {\mu }}-\partial A_{\mu }^{\prime }/\partial x^{\prime {%
\nu }},$ with $H$ being the corresponding eigenvalue.

\subsection{Equations of motion}

Equations of motion, or the relationships between differentials tangent to
the particle orbit, can be obtained as Euler equations of the transformed
variational principle (\ref{summary}). Let us find the first variation of $S$
assuming $(x^{\prime \alpha },\phi ,\hat{\mu},{u_{\parallel }})$ to be
independent:
\begin{multline}
\delta S=\int \left( -qF_{\mu \nu }+d_{\nu \mu }\right) \delta x^{\prime \nu
}\mathrm{d}{x^{\prime \mu }}+  \notag \\
+\left( \frac{q\tau _{\mu }}{u_{0}}\frac{\partial \left( H^{\ast }\hat{\mu}%
\right) }{\partial \hat{\mu}}+q\frac{\partial \bar{a}_{\mu }}{\partial \hat{%
\mu}}+\frac{\chi _{\mu }}{2}\right) \left[ \delta \hat{\mu}\mathrm{d}{%
x^{\prime \mu }-}\mathrm{d}\hat{\mu}\delta {x^{\prime \mu }}\right] {+} \\
{+}\left( l_{\mu }+\frac{u{_{\parallel }}\tau _{\mu }}{u_{0}}+s_{\bot \mu
}\right) \left[ \delta u{_{\parallel }}\mathrm{d}{x^{\prime \mu }}-\mathrm{d}%
u{_{\parallel }\delta x^{\prime \mu }}\right] +\delta \hat{\mu}\mathrm{d}%
\phi -\mathrm{d}\hat{\mu}\delta \phi .  \notag
\end{multline}%
Here we dropped bars over $u{_{\parallel },}u_{0}$ since all oscillating
corrections are left behind, and introduced a new tensor notation
\begin{equation}
d_{\nu \mu }\equiv \mathrm{Rot}_{\nu }\left[ u{_{\parallel }}l_{\mu
}+u_{0}\tau _{\mu }+q\bar{a}_{\mu }+\bar{u}{_{\bot \mu }}+\frac{1}{2}\hat{\mu%
}\chi _{\mu }\right] ,
\end{equation}%
where the operator $\mathrm{Rot}_{\nu }$ is defined by $\mathrm{Rot}_{\nu
}f_{\mu }\equiv \partial f_{\mu }/\partial x^{\prime \nu }-\partial f_{\nu
}/\partial x^{\prime \mu }.$ Furthermore, the partial derivative of the
wave-induced drift is denoted as%
\begin{equation*}
s_{\bot \mu }\equiv \frac{\partial \bar{u}{_{\bot \mu }}}{\partial u{%
_{\parallel }}}=\frac{1}{2H}b_{\mu }^{\;\lambda }\bar{f}_{\lambda \nu
}l^{\nu }.
\end{equation*}

Extrema of the functional are achieved for world lines where $\delta S=0$
for all variations of independent variables. This yields the Euler equations
as
\begin{eqnarray}
\mathrm{d}\hat{\mu} &=&0,  \label{muconserve} \\
\left( \frac{q\tau _{\mu }}{u_{0}}\frac{\partial \left( H^{\ast }\hat{\mu}%
\right) }{\partial \hat{\mu}}+q\frac{\partial \bar{a}_{\mu }}{\partial \hat{%
\mu}}+\frac{\chi _{\mu }}{2}\right) \mathrm{d}{x^{\prime \mu }+}\mathrm{d}%
\phi &=&0,  \label{dphi} \\
\left( l_{\mu }+\frac{u{_{\parallel }}\tau _{\mu }}{u_{0}}+s_{\bot \mu
}\right) \mathrm{d}{x^{\prime \mu }} &=&{0,}  \label{scal}
\end{eqnarray}%
\begin{equation}
\left( -qF_{\mu \nu }+d_{\nu \mu }\right) \mathrm{d}{x^{\prime \mu }}-\left(
l_{\nu }+\frac{u{_{\parallel }}\tau _{\nu }}{u_{0}}+s_{\bot \nu }\right)
\mathrm{d}u{_{\parallel }}=0.  \label{motion}
\end{equation}%
The first equation here confirms that $\widehat{\mu }$ is an integral of
motion, the second equation determines the rate of rotation along the Larmor
orbit and is not needed for our purposes here. By multiplying the 4-vector
equation (\ref{motion}) by $\mathrm{d}{x^{\prime \nu }}$ and taking the sum
in $\nu $ one can recover the third scalar equation (\ref{scal}) multiplied
by $\mathrm{d}u{_{\parallel }}$, i.e. out of five equations for four unknown
functions $\mathrm{d}{x^{\prime \mu }/}\mathrm{d}u{_{\parallel }}$ only four
are independent, as it should be for solvability. This allows to determine
the non-trivial solutions (at least formally) and use them as coefficients
of the gyrokinetic equation.

It is convenient to rewrite the 4-vector equation (\ref{motion}) through its
projections on the orthogonal vectors of the basis $(\mathbf{\tau
,l,l^{\prime },l^{\prime \prime }}).$ Taking scalar products with $\left(
\mathbf{l^{\prime },l^{\prime \prime }}\right) $ we get
\begin{eqnarray}
\left( qHl_{\mu }^{\prime \prime }+l^{\prime \nu }d_{\nu \mu }\right)
\mathrm{d}{x^{\prime \mu }+}s_{\bot \nu }l^{\prime \nu }\mathrm{d}u{%
_{\parallel }} &=&{0,}  \label{first} \\
\left( -qHl_{\mu }^{\prime }+l^{\prime \prime \nu }d_{\nu \mu }\right)
\mathrm{d}{x^{\prime \mu }+}s_{\bot \nu }l^{\prime \prime \nu }\mathrm{d}u{%
_{\parallel }} &=&{0.}  \label{sec}
\end{eqnarray}%
These two equations determine the drift velocity. Taking the scalar product
with $\mathbf{l,}$ we recover the following equation governing acceleration
parallel to the magnetic field
\begin{equation}
\mathrm{d}u{_{\parallel }-}qE\tau _{\mu }\mathrm{d}{x^{\prime \mu }}+l^{\nu
}d_{\nu \mu }\mathrm{d}{x^{\prime \mu }}=0.  \label{upa}
\end{equation}%
Then, the last equation looks like
\begin{equation}
u{_{\parallel }}\mathrm{d}u{_{\parallel }+}u_{0}qEl_{\mu }\mathrm{d}{%
x^{\prime \mu }}-u_{0}\tau ^{\nu }d_{\nu \mu }\mathrm{d}{x^{\prime \mu }}=0,
\label{fourth}
\end{equation}%
which is the energy conservation law. However, as shown above, instead one
can use another, equivalent but much more compact equation (\ref{scal}). It
describes the relationship between $u{_{\parallel }}$\ and the differentials
of the guiding-center position
\begin{equation}
\left( u_{0}l_{\mu }+u{_{\parallel }}\tau _{\mu }+u_{o}s_{\bot \mu }\right)
\mathrm{d}{x^{\prime \mu }}=0.  \label{udx}
\end{equation}

Equations (\ref{first}),(\ref{sec}) can be rewritten to exclude $\mathrm{d}u{%
_{\parallel }}$ using Eq.(\ref{upa}):%
\begin{eqnarray}
\left( qHl_{\mu }^{\prime \prime }+l^{\prime \nu }d_{\nu \mu }+s_{\bot \nu
}l^{\prime \nu }\left[ qE\tau _{\mu }-l^{\nu }d_{\nu \mu }\right] \right)
\mathrm{d}{x^{\prime \mu }} &=&0,  \label{first1} \\
\left( -qHl_{\mu }^{\prime }+l^{\prime \prime \nu }d_{\nu \mu }+s_{\bot \nu
}l^{\prime \prime \nu }\left[ qE\tau _{\mu }-l^{\nu }d_{\nu \mu }\right]
\right) \mathrm{d}{x^{\prime \mu }} &=&{0.}  \label{sec1}
\end{eqnarray}

It is possible to formally solve equations (\ref{first})-(\ref{udx}) in
terms of the basis vectors. To do this we introduce new notations,%
\begin{eqnarray}
v_{\mu } &=&u_{o}l_{\mu }+u{_{\parallel }}\tau _{\mu }, \\
\bar{u}^{\mu } &=&u_{\parallel }l_{\mu }+u{_{0}}\tau _{\mu },
\end{eqnarray}%
and
\begin{equation}
d_{pq}\equiv p^{\nu }d_{\nu \mu }q^{\mu },\qquad s_{p}=s_{\bot \nu }p^{\nu },
\end{equation}
for arbitrary four-vectors $\mathbf{p,q}$.\ Note that $v_{\mu }\bar{u}^{\mu
}=0$ and these vectors can be used in place of $l_{\mu },\tau _{\mu }.$

Searching for the solution in the form $\mathrm{d}{x^{\prime \mu }=}\bar{u}%
^{\mu }+al^{\prime \mu }+bl^{\prime \prime \mu }+cv_{\mu }$, and solving
equations (\ref{udx}),(\ref{first1}),(\ref{sec1}) with respect to $a,b,c$ up
to the first order in $1/qH$, we get%
\begin{equation}
\mathrm{d}{x^{\prime \mu }=}\left( \bar{u}^{\mu }+\frac{1}{qH}\left[
l^{\prime \prime \mu }d_{l^{\prime }u}-l^{\prime \mu }d_{l^{\prime \prime
}u}+\frac{u_{o}v^{\mu }}{u_{o}^{2}-u_{\parallel }^{2}}\left( s_{l^{\prime
\prime }}d_{l^{\prime }u}-s_{l^{\prime }}d_{l^{\prime \prime }u}\right) %
\right] \right) \mathrm{d}s^{\prime },  \label{soldx}
\end{equation}%
where $\mathrm{d}s^{\prime }$ is an arbitrary scalar function. Similarly,
from Eq.(\ref{upa}) we find%
\begin{eqnarray}
\mathrm{d}u{_{\parallel }} &{=}&\left( qE-d_{l\tau }\right) u_{o}\mathrm{d}%
s^{\prime }-  \label{soldu} \\
&&-\frac{1}{qH}\left[ d_{ll^{\prime \prime }}d_{l^{\prime }u}-d_{ll^{\prime
}}d_{l^{\prime \prime }u}-\frac{u_{o}u_{\parallel }}{u_{o}^{2}-u_{\parallel
}^{2}}\left( qE-d_{l\tau }\right) \left( s_{l^{\prime \prime }}d_{l^{\prime
}u}-s_{l^{\prime }}d_{l^{\prime \prime }u}\right) \right] \mathrm{d}%
s^{\prime }.
\end{eqnarray}%
The zero-order terms describe the particle trajectory in quasi-uniform
fields. In the non-relativistic limit the $d_{l\tau }$ term in Eq.(\ref%
{soldu}) is responsible for the parallel diamagnetic force $\sim \mu \nabla
B,$ while $\bar{u}^{\mu }$ contains the parallel velocity as well as the $%
E\times B$ -drift. The first order corrections (in square brackets) describe
drifts due to spatial and temporal inhomogeneities.

Note that the equations of motion in the present formulation result
different and significantly more complicated than in the previous
formulation.\cite{BT2} Thus, it should be used only if the $E=0$ limit is
expected to occur in conjunction with the short wavelength wave fields, in
which case the previous formulation fails.

In Section V we will need to relate the gyrokinetic displacements to the
particle displacement, i.e., express $\mathrm{d}s^{\prime }$ in terms of the
particle displacement $\mathrm{d}s.$ Below we present a proof that the two
displacements coincide in the zero-order approximation.

According to the definitions,%
\begin{equation*}
\mathrm{d}s^{2}=g_{\mu \nu }\mathrm{d}x^{\mu }\mathrm{d}x^{\nu }\approx
g_{\mu \nu }(\mathbf{x}^{\prime })\left[ \mathrm{d}x^{\prime \mu }+\frac{%
\partial r_{1}^{\mu }}{\partial \phi }\mathrm{d}\phi \right] \left[ \mathrm{d%
}x^{\prime \nu }+\frac{\partial r_{1}^{\nu }}{\partial \phi }\mathrm{d}\phi %
\right] .
\end{equation*}%
This can be expanded to yield%
\begin{equation*}
\mathrm{d}s^{2}\approx g_{\mu \nu }^{\prime }\mathrm{d}x^{\prime \mu }%
\mathrm{d}x^{\prime \nu }+2g_{\mu \nu }^{\prime }\mathrm{d}x^{\prime \nu }%
\frac{\partial r_{1}^{\mu }}{\partial \phi }\mathrm{d}\phi +\frac{\partial
r_{1}^{\mu }}{\partial \phi }\frac{\partial r_{1\mu }}{\partial \phi }%
\mathrm{d}\phi ^{2}.
\end{equation*}%
The differentials in the right-hand side are supposed to be taken along the
particle trajectory, i.e., should satisfy the equations of motion, (\ref%
{dphi}),(\ref{soldx}). Using them in zero order,%
\begin{align}
u_{o}^{\prime }\mathrm{d}\phi +qH\tau _{\mu }\mathrm{d}x^{\prime \mu }& =0,
\\
\mathrm{d}x^{\prime \mu }-\bar{u}^{\prime \mu }\mathrm{d}s^{\prime }& =0,
\notag
\end{align}%
we obtain%
\begin{equation*}
\mathrm{d}s^{2}=\bar{u}^{\prime \mu }\bar{u}_{\mu }^{\prime }\mathrm{d}%
s^{\prime 2}+\frac{\partial r_{1}^{\mu }}{\partial \phi }\frac{\partial
r_{1\mu }}{\partial \phi }q^{2}H^{2}\mathrm{d}s^{\prime 2}.
\end{equation*}%
Furthermore, available expressions for $r_{1}^{\mu }$ can be used to yield%
\begin{equation*}
\frac{\partial r_{1}^{\mu }}{\partial \phi }\frac{\partial r_{1\mu }}{%
\partial \phi }=-\frac{w^{2}}{q^{2}H^{2}},
\end{equation*}%
so that, finally,%
\begin{equation}
\mathrm{d}s=\mathrm{d}s^{\prime }.  \label{dsds'}
\end{equation}

\section{Equivalence of gyrokinetic theories}

The gyrokinetic transformation is about finding and using the adiabatic
invariant, which corresponds to the fast Larmor rotation, and in the
gyrokinetic limit becomes the first integral of motion. As such it splits
the phase space into a sequence of hypersurfaces. The invariant
hypersurfaces should be the same for all gyrokinetic theories, unless they
invoke an additional first integral (adiabatic invariant) of some sort.
Thus, we conclude, that in the same phase space the adiabatic invariants of
different gyrokinetic theories should be in functional dependence. There is
a catch, however. Indeed, the phase space of a canonical theory is not
gauge-invariant, since the canonical momenta depend on the gauge of the
electromagnetic field. Thus, the adiabatic invariants with different gauges
can (and even should) be different. For a non-canonical Lagrangian theory
the phase space is chosen in an observable way, but this does not preclude
similar ambiguity. Indeed, the gauge invariance of the Lagrangian itself
allows for different expressions to appear.

There is another sort of ambiguity in terms related to division of the field
into the \textquotedblleft background\textquotedblright\ and the
\textquotedblleft wave\textquotedblright\ parts. The use of the two
expansion parameters, $\varepsilon $ and $\lambda $, instead of a single one
is due to convenience, rather than necessity. It is especially evident in
our covariant treatment, where all artificial restrictions on the wave-field
introduced in conventional theories are removed. Indeed, both parameters are
inversely proportional to the background magnetic field, and the two
expansions can be effectively replaced by a single one by assuming that
\emph{all }nonuniformity of fields is due to effective \textquotedblleft
waves\textquotedblright , i.e., by representing the background field via its
Fourier expansion rather than its Taylor expansion. Obviously, if the
variation of the background field is small on the Larmor scale, i.e., $%
\varepsilon \ll 1$, then the deviation from the unperturbed orbit will also
be small, i.e., $\lambda \ll 1$, and the applicability of the wave-field
expansion for the variable part of the background field is ensured. Unlike
the Taylor expansion, Fourier expansion is non-local, and as such has
potentially better convergence for finite values of the deviation. This is
very important for the gyrokinetic theory, which is used primarily for
description of effects due to the finite Larmor radius of particles.
Unfortunately, non-local theories are far less convenient from the
mathematical viewpoint.

The two ambiguities described above are more-or-less controlled by the user,
so that they can be counted among advantages rather than difficulties of the
theory. If the adiabatic invariant is the only integral of motion, and the
gyrophase is defined as its canonically conjugate angle variable, the
freedom of variation becomes confined to different transformations within
the subspace of the rest of the phase variables. This can be defined as a
class of equivalence of different gyrokinetic theories:

\emph{If, under the same gauge and the same division of the wave-field, the
adiabatic invariants of two theories are in functional dependence, then the
theories are \textquotedblleft equivalent\textquotedblright ,}

i.e., the definitions of the gyrophase coincide, and all the difference is
just in possible transformations of the \textquotedblleft
drift\textquotedblright\ variables. As stressed above, theories should be
equivalent if there are no more integrals of motion that could be expressed
via local parameters. (Global integrals of motion are here irrelevant, if
the derivation procedure is local as usual.)

The above argument holds for any particular power of expansion in $%
\varepsilon .$ However, the definition of the adiabatic invariant of order $%
n $ may depend on the existence of other approximate integrals of motion of
order $n,$ since any function of all integrals of motion is the integral of
motion and there is no formal means to distinguish between them. In
particular, in the first order the electromagnetic field is uniform, and in
theories with slow drifts even the electric field is zero, so that there are
at least two more integrals of motion, such as the parallel velocity and the
perpendicular gyrocenter position. As a result, \emph{in the first order},
the definition of the \textquotedblleft adiabatic
invariant\textquotedblright\ is very slack, and besides the gauge, allows
for two more arbitrary functions to enter the definition. This fact is
reflected in our paper,\cite{BT} where it has been shown that the
gyrokinetic transformation in the first order is defined with six free
parameters.

For higher-order theories this is dangerous, because by choosing one of the
definitions at an early stage, one can be left without any solution in the
next order, where the other approximate integrals of motion may disappear.
Thus, the solubility condition for the second order approximation removed
all freedom from the definition of the adiabatic invariant in our subsequent
papers,\cite{BT1,BT2} (though left some in other respects.) Another
possibility is the survival of additional integral(s) of motion in the next
order due to some ordering or symmetry assumptions. For example, all
previous (except \cite{BT1,BT2}) gyrokinetic theories assume that the
parallel electric field and the background field inhomogeneity are zero, and
the parallel wavelength of the perturbation is infinite, while describing
the effect of the wave. This means, that at least the parallel momentum of
the particle is retained as an approximate integral of motion in this order,
and there is a possible ambiguity in the definition of the \textquotedblleft
adiabatic invariant\textquotedblright . (Of course, nobody is so dumb as to
insert the parallel momentum in the definition by hand. Still, while using
multi-step approximations, it is necessary to keep the possible ambiguity in
mind.) At the moment we do not know of any danger related to this
possibility, and it seems that, though the adiabatic invariant can be
potentially defined in different ways, it is impossible to say which one is
\textquotedblleft right\textquotedblright . Thus, the gyrokinetic theories
are potentially non-equivalent. In particular, different choices of free
parameters in Ref.\cite{BT} lead to non-equivalent theories, when the
corresponding definitions of the magnetic moment are different.

So, why is this \textquotedblleft non-equivalence\textquotedblright\
important? The reason lies in the internal machinery of the derivation. If
the gyrokinetic transformations are non-equivalent, the definitions of the
gyrophase are different, and, hence, so are the definitions of the
gyrocenter/guiding center. As a consequence, the \textquotedblleft drift
trajectories\textquotedblright\ in different representations will differ by
a distance of the order of the Larmor radius ($\times \varepsilon ^{n-1}$),
depending on the gyrophase. Drift trajectories are routinely used for
confinement analysis. If they form a drift surface, the conclusion is that
the confinement is achieved. Now we see that this logic works in only one
direction. The reverse is not true: even if the drift trajectories form a
surface in one representation, it will not necessarily be the same one you
are using!

\emph{If the gyrokinetic theories are not equivalent, they have close but
essentially different drift trajectories.}

\section{Cyclotron damping}

A fundamental issue of for the kinetic description of relativistic
magnetized plasmas is the consistent treatment of cyclotron losses. In this
section we describe the radiation losses of a charged particle accelerated
by an external electromagnetic field as a single-particle classical process.
However, one should keep in mind that in cold, dense plasmas there may be
interference of cyclotron radiation from different particles, thus modifying
both the outcoming radiation and its net effect on the particle. In this
limit the \textquotedblleft cyclotron losses\textquotedblright\ are better
described by means of collision integrals and plasma waves.

Radiation damping of the motion of a particle in an electromagnetic field is
given by the force \cite{LL3}%
\begin{equation*}
f_{i}=\frac{2q_{a}^{2}}{3c}\left( \frac{\mathrm{d}^{2}u_{i}}{\mathrm{d}s^{2}}%
-u_{i}u^{k}\frac{\mathrm{d}^{2}u_{k}}{\mathrm{d}s^{2}}\right) .
\end{equation*}%
Normally it should be a small correction to the electromagnetic force, but
can also be large for ultra-relativistic particles.

This expression can be translated to the covariant form by substitution of
covariant derivatives in place of $\mathrm{d}^{2}u_{i}/\mathrm{d}s^{2}$ .
Using the equations of motion, we get%
\begin{equation*}
\frac{\mathrm{D}^{2}u_{i}}{\mathrm{d}s^{2}}=q^{2}F_{ik}F^{kl}u_{l}+q\frac{%
\mathrm{D}F_{ik}}{\partial x^{l}}u^{l}u^{k}.
\end{equation*}%
Substituting it into the expression for the force, we can deduce the
four-acceleration caused by radiation as
\begin{equation}
\left( \frac{\mathrm{D}u_{i}}{\mathrm{d}s}\right) _{r}=\frac{2}{3}r_{a}q%
\left[ qF_{ik}F^{kl}u_{l}+qu_{i}(F_{kl}u^{l})^{2}+\frac{\mathrm{D}F_{ik}}{%
\partial x_{l}}u^{l}u^{k}\right] ,  \label{cyfi}
\end{equation}%
where $r_{a}=q_{a}^{2}/m_{a}c^{2}$ is the electromagnetic \textquotedblleft
radius\textquotedblright\ of the particle, and $q=q_{a}/m_{a}c^{2}$ is the
normalized charge.

From the normalization it is obvious, that the order of the correction
caused by the damping is $\varepsilon ^{-1}r_{a}/\rho _{L},$ which is,
formally, $\varepsilon ^{-2}.$ However, the ratio $r_{a}/\rho _{L}$ is very
small in realistic situations, so that we shall assume it sufficient to
calculate the trajectory ($u_{l}(s)$) just in the first order in the
Larmor-radius expansion. In this order the electromagnetic field, as well as
the coordinate system are constant, i.e., the last term in the expression (%
\ref{cyfi}) can be neglected. Thus%
\begin{equation}
\left( \mathrm{D}u_{\mu }\right) _{r}=\frac{2}{3}r_{a}q^{2}\left[ F_{\mu \nu
}F^{\nu \lambda }u_{\lambda }-(u^{\alpha }F_{\alpha \nu }F^{\nu \lambda
}u_{\lambda })u_{\mu }\right] \mathrm{d}s.  \label{cyfi1}
\end{equation}

This expression has a general form of acceleration due to friction versus
some medium with characteristic velocity $\mathbf{u}_{0}=\mathrm{F}^{2}%
\mathbf{u}$:

\begin{equation}
\mathbf{a}=\alpha _{r}\left[ \mathbf{u}_{0}\mathrm{-}\mathbf{u}\left(
\mathbf{uu}_{0}\right) \right] .
\end{equation}%
It is consistent with (\ref{cyfi1}) for $\alpha _{r}=2r_{a}q^{2}/3,$ and $%
\mathbf{a}\mathrm{=D}\mathbf{u}\mathrm{/d}s\mathrm{.}$ It includes
deceleration due to both, the cyclotron/synchrotron radiation, which is
proportional to $H^{2}$, and the bremsstrahlung, which is proportional to $%
E^{2}.$

Note also, that the acceleration due to the mean field is proportional to $%
\mathrm{F}\mathbf{u}\mathrm{,}$ with $\mathrm{F}$ being antisymmetric, while
the radiative deceleration is proportional to $\mathrm{T}\mathbf{u}$ , with
a symmetric $\mathrm{T}$, but such that $\mathbf{u}\mathrm{T}\mathbf{u}%
\mathrm{=0.}$

Next step is to rewrite the friction in terms of the gyrokinetic variables.
Using the representations of the world-velocity,
\begin{equation}
u_{\nu }=\bar{u}_{\nu }+w\left( l_{\nu }^{\prime }\cos \phi +l_{\nu
}^{\prime \prime }\sin \phi \right) \equiv u_{\Vert \nu }+u_{\bot \nu },
\label{un11}
\end{equation}%
and of the Faraday tensor, (\ref{Ft}), we find%
\begin{equation*}
\mathbf{a}=\alpha _{r}\left[ H^{2}\mathrm{bb}\mathbf{u}_{\bot }+E^{2}\mathrm{%
cc}\mathbf{u}_{\Vert }-\mathbf{u}\left( H^{2}\mathbf{u}_{\bot }\mathrm{bb}%
\mathbf{u}_{\bot }+E^{2}\mathbf{u}_{\Vert }\mathrm{cc}\mathbf{u}_{\Vert
}\right) \right] .
\end{equation*}%
Using definitions of $\mathrm{b}$ and $\mathrm{c}$, it can be further
transformed to%
\begin{equation*}
\mathbf{a}=\alpha _{r}\left[ -H^{2}\mathbf{u}_{\bot }+E^{2}\mathbf{u}_{\Vert
}-\mathbf{u}\left( w^{2}H^{2}+(1+w^{2})E^{2}\right) \right] ,
\end{equation*}%
or%
\begin{equation*}
\mathbf{a}=-\alpha _{r}\left( H^{2}+E^{2}\right) \left[ (1+w^{2})\mathbf{u}%
_{\bot }+w^{2}\mathbf{u}_{\Vert }\right] .
\end{equation*}%
Since $\mathbf{u}_{\bot }^{2}=-w^{2}$ and $\mathbf{u}_{\Vert }^{2}=1+w^{2}$
it is obvious that $\mathbf{au}\mathrm{=0}$ as desired .

The cyclotron friction is a small effect, so it can be evaluated in the
first-order gyrokinetic approximation, i.e., with all fields being uniform.
Also, we can assume components of the metric tensor to be constant in this
approximation. In this way we neglect accelerations due to gravity and the
curvature of field lines as compared to the acceleration on the Larmor
orbit. Then, the equation
\begin{equation*}
\mathrm{D}\mathbf{u}_{r}\mathrm{=}\mathbf{a}\mathrm{d}s
\end{equation*}%
can be rewritten as%
\begin{multline}
\left[ \left( l_{\nu }^{\prime }\cos \phi +l_{\nu }^{\prime \prime }\sin
\phi \right) +\frac{w}{u_{o}}\tau _{\mu }\right] \mathrm{d}w_{r}+ \\
+\left[ l_{\mu }+\frac{{u_{\parallel }}}{u_{o}}\tau _{\mu }\right] \mathrm{d}%
{u_{\parallel r}}+\left( -l_{\nu }^{\prime }\sin \phi +l_{\nu }^{\prime
\prime }\cos \phi \right) w\mathrm{d}{\phi }_{r}= \\
=-\alpha _{r}\left( H^{2}+E^{2}\right) \left[ w(1+w^{2})\left( l_{\nu
}^{\prime }\cos \phi +l_{\nu }^{\prime \prime }\sin \phi \right)
+w^{2}\left( {u_{\parallel }}l_{\mu }+u_{o}\tau _{\mu }\right) \right]
\mathrm{d}s.  \notag
\end{multline}%
Taking scalar products with the basis vectors we find%
\begin{equation}
\left\{
\begin{array}{l}
\mathrm{d}{\phi }_{r}=0, \\
\mathrm{d}w_{r}=-\frac{2}{3}r_{a}q^{2}\left( H^{2}+E^{2}\right) w(1+w^{2})%
\mathrm{d}s, \\
\mathrm{d}{u_{\parallel r}=}-\frac{2}{3}r_{a}q^{2}\left( H^{2}+E^{2}\right)
w^{2}{u_{\parallel }}\mathrm{d}s.%
\end{array}%
\right.  \label{dfdwdu}
\end{equation}%
Here the differentials $\mathrm{d}s$ can be identified with $\mathrm{d}%
s^{\prime }$, which was used in Section III for description of the
gyrokinetic trajectory, due to identity (\ref{dsds'}). As a result, it is
easy to \textquotedblleft tweak\textquotedblright\ the equations of motion
to include the radiation damping as a small correction:%
\begin{equation}
\left\{
\begin{array}{l}
\mathrm{d}x^{\prime \mu }=\mathrm{d}x_{g}^{\prime \mu }+0 \\
\mathrm{d}{u_{\parallel }=}\mathrm{d}{u_{\parallel g}+}\mathrm{d}{%
u_{\parallel r}} \\
\mathrm{d}\hat{\mu}=0+w\mathrm{d}w_{r}/qH \\
\mathrm{d}{\phi =}\mathrm{d}{\phi }_{g}{+}\mathrm{d}{\phi }_{r}%
\end{array}%
\right. ,  \label{traject}
\end{equation}%
where $\mathrm{d}x_{g}^{\prime \mu }\mathrm{d}{u_{\parallel g}}\mathrm{d}{%
\phi }_{g}$ describe the gyrokinetic trajectory, while $\mathrm{d}{%
u_{\parallel r},}\mathrm{d}w_{r},\mathrm{d}{\phi }_{r}$ are corrections due
to the cyclotron damping. All these differentials along the particle orbit
are already found in Section III and above, and are ready to use.

\section{Kinetic equation}

If equations of motion contain dissipative terms, i.e., the particle energy
and momentum are lost along the trajectory, the form of the kinetic equation
itself may change. It is necessary to check the Liouville's theorem, and
introduce corrections if the phase volume is not conserved.

First, let us define the phase space as a direct product of the space-time
by the four-velocity space. Note, that scalar products are defined in each
subspace, and \emph{between} vectors of different subspaces, but not in the
phase-space as a whole - you cannot add velocities to distances. However,
all trajectories or world-lines of particles, can be parameterized by the
proper time $s$, whose definition involves only the space-time subspace. In
effect, zero length is ascribed to displacements in velocity directions.

Coordinates in the phase-space are introduced in accordance with existing
scalar products. Namely, initially, for \emph{both} subspaces we use the
same systems of basis vectors, i.e., for an 8-dimensional space we use two
copies of four-dimensional basis vectors of space-time.

Consider a flux-tube of trajectories. At a given point $\mathbf{(x,u)}$ it
could be characterized by the direction of its central line,$\mathrm{(d}%
\mathbf{x}\mathrm{\mathbf{,}D}\mathbf{u}\mathrm{)/d}s$, its cross-section, $%
\mathrm{\delta }V$, and the flux density, $f\mathrm{(}\mathbf{x,u}\mathrm{)}$%
, which is identified as the distribution function of particles. However,
the definitions of the flux density and the cross-section area are subject
to the definition of the scalar product, which is not fully defined.
Ascribing zero length to velocity displacements, as in the standard
parameterization, means that the direction of the central line is in effect $%
\mathrm{(d}\mathbf{x}\mathrm{,D}\mathbf{u}\mathrm{)/d}s\mathrm{\rightarrow (}%
\mathbf{u}\mathrm{,}\mathbf{0}\mathrm{)}$. Then the phase-space volume $%
\mathrm{\delta }V$ can be defined as a product of the volume element in the
space-time, $\mathrm{\delta }V_{x}$, which is orthogonal to $\mathbf{u}$,
and the element of the total velocity subspace, $\mathrm{\delta }V_{u}$, so
that $\mathrm{\delta }V=\mathrm{\delta }V_{x}\mathrm{\delta }V_{u}$. The
volume element in the velocity space, $\mathrm{\delta }V_{u}$, is also
orthogonal to $\mathbf{u}$, since the motion occurs on the hypersurface $%
\mathbf{u}^{2}=1$, which is orthogonal to its radius, $\mathbf{u}\mathrm{.}$

The flux of trajectories is continuous, i.e., the flux lines do not end, if
the particles do not disappear. This means that $f\mathrm{(}\mathbf{x}%
\mathrm{,}\mathbf{u}\mathrm{)\delta }V=const$, i.e.,%
\begin{equation*}
\mathrm{d}\left( f\mathrm{(}\mathbf{x}\mathrm{,}\mathbf{u}\mathrm{)\delta }%
V\right) =0,
\end{equation*}%
which is a general form of the kinetic equation. If, additionally, the
Liouville's theorem holds, i.e.,%
\begin{equation*}
\mathrm{d\delta }V=0,
\end{equation*}%
then the kinetic equation obtains its classic form,%
\begin{equation*}
\mathrm{d}f\mathrm{(}\mathbf{x}\mathrm{,}\mathbf{u}\mathrm{)}=0.
\end{equation*}

Note, that the differential $\mathrm{d}$ takes place in the 7-dimensional
phase-space, while the volume element $\mathrm{\delta }V$ is 6-dimensional,
as it describes an element of the hypersurface. Consequently, $f$ $\mathrm{%
\delta }V$ is the density of particles in their rest-frame in one particular
moment. Defined in this way, the phase-space volume and the distribution
function coincide with their non-relativistic analogs in the corresponding
limit.

The full volume element in the space-time is $\mathrm{\delta }V_{\sigma
}=\epsilon _{\alpha \beta \gamma \delta }\mathrm{\delta }x_{1}^{\alpha }%
\mathrm{\delta }x_{2}^{\beta }\mathrm{\delta }x_{3}^{\gamma }\mathrm{\delta }%
x_{4}^{\delta },$ while its sub-volume (section area) orthogonal to the unit
vector $\mathbf{u}$ is
\begin{equation*}
\mathrm{\delta }V_{x}=\epsilon _{\alpha \beta \gamma \delta }u^{\alpha }%
\left[ \mathrm{\delta }x_{1}^{\beta }\mathrm{\delta }x_{2}^{\gamma }\mathrm{%
\delta }x_{3}^{\delta }+\mathrm{\delta }x_{0}^{\beta }\mathrm{\delta }%
x_{1}^{\gamma }\mathrm{\delta }x_{3}^{\delta }-\mathrm{\delta }x_{0}^{\beta }%
\mathrm{\delta }x_{1}^{\gamma }\mathrm{\delta }x_{2}^{\delta }-\mathrm{%
\delta }x_{0}^{\beta }\mathrm{\delta }x_{2}^{\gamma }\mathrm{\delta }%
x_{3}^{\delta }\right] .
\end{equation*}%
Here $\mathrm{\delta }x_{i}^{\beta }$ is the $\beta $-component of the $i$%
-th box vector, and $\epsilon _{\alpha \beta \gamma \delta }$ is the
absolutely antisymmetric tensor. Thus, the volume elements are scalars, so
that if they are constant in one coordinate system, they are constant in all
others. In particular, we can choose the local co-moving reference frame, in
which $u^{\alpha }=0$ and $u_{\alpha }=0$ for all $\alpha \neq 0$ (this
requires also $g_{\mu 0}=0$). Then%
\begin{equation}
\mathrm{\delta }V_{x}=\epsilon _{0123}u^{0}\mathrm{\delta }x^{1}\mathrm{%
\delta }x^{2}\mathrm{\delta }x^{3}=u^{0}\sqrt{-g}\mathrm{\delta }x^{1}%
\mathrm{\delta }x^{2}\mathrm{\delta }x^{3},  \label{dvx}
\end{equation}%
where the box sides are taken along the basis vectors. In words the
invariant space subvolume is defined as follows: if there is a particle with
velocity $\mathbf{u,}$ go into its rest frame, and put it in a box. This is
a unique way of measurement, i.e., a scalar.

The effective volume element in the velocity-subspace we define as
\begin{equation}
\mathrm{\delta }V_{u}=-\epsilon _{\mu \nu \lambda \kappa }u^{\mu }\mathrm{%
\delta }u_{1}^{\nu }\mathrm{\delta }u_{2}^{\lambda }\mathrm{\delta }%
u_{3}^{\kappa }=-\epsilon ^{\mu \nu \lambda \kappa }u_{\mu }\mathrm{\delta }%
u_{1\nu }\mathrm{\delta }u_{2\lambda }\mathrm{\delta }u_{3\kappa }.
\label{dvu0}
\end{equation}%
Its definition is important for calculating moments of the distribution
function, such as the four-current density. We have defined it as a scalar,
i.e., an invariant under Lorentz transformations, and it should coincide
with corresponding definitions by other authors. Indeed, in the co-moving
frame it becomes%
\begin{equation}
\mathrm{\delta }V_{u}=-\epsilon ^{0123}u_{0}\mathrm{\delta }u_{1}\mathrm{%
\delta }u_{2}\mathrm{\delta }u_{3}=\frac{\mathrm{\delta }u_{1}\mathrm{\delta
}u_{2}\mathrm{\delta }u_{3}}{u^{0}\sqrt{-g}},  \label{dvu}
\end{equation}%
which is listed as an invariant velocity volume in Refs.\cite{Misner,LL1} It
can also be expressed via the $\delta $-function as%
\begin{equation}
\mathrm{\delta }V_{u}=\delta \left( 1-\sqrt{u_{\mu }u^{\mu }}\right) \mathrm{%
d}^{4}\mathbf{u}=\delta \left( 1-\sqrt{u_{\alpha }u^{\alpha }}\right)
\epsilon ^{\mu \nu \lambda \kappa }\mathrm{\delta }u_{0\mu }\mathrm{\delta }%
u_{1\nu }\mathrm{\delta }u_{2\lambda }\mathrm{\delta }u_{3\kappa }.
\label{dvu1}
\end{equation}%
Indeed, this is also a scalar expression, and in the co-moving frame it
obviously coincides with Eq.(\ref{dvu}).

Multiplying the volume elements (\ref{dvx}) and (\ref{dvu}) we get a simple
expression for the phase-space volume in the co-moving frame%
\begin{equation}
\mathrm{\delta }V=\mathrm{\delta }x^{1}\mathrm{\delta }x^{2}\mathrm{\delta }%
x^{3}\mathrm{\delta }u_{1}\mathrm{\delta }u_{2}\mathrm{\delta }u_{3}.
\label{dv}
\end{equation}%
Here the box sides $\mathrm{\delta }x^{\alpha },\mathrm{\delta }u_{\alpha }$
are defined along the corresponding basis vectors, i.e., are pure-space
coordinates and velocities.

The next question is how this box changes along the particle trajectory. To
describe it, first note, that each direction in space, $\mathbf{n}$,
specifies a rectangle cross-section of the box, $[\mathbf{x}_{\mathbf{n}},%
\mathbf{u}_{\mathbf{n}}].$ Such cross-section is initially a rectangle,
since the volume is defined by a simple product of the velocity and space
boxes. Consider a small evolutionary change of the box, and, hence, of the
cross-section. The deformation of the rectangle is small and is due to
different velocities (in $\mathbf{x}$) and different accelerations (in $%
\mathbf{u}$) of particles. Since it is small we can consider it linear, and
as such it is a superposition of deformations proportional to $\mathbf{u}%
\mathrm{d}s$ and to $\mathbf{a}\mathrm{d}s=\mathrm{D}\mathbf{u}\mathrm{.}$

The velocity-deformation, $x_{\mathbf{n}}^{\prime }=x_{\mathbf{n}}\mathrm{+}%
u_{\mathbf{n}}\mathrm{d}s,$ conserves the area, since it acts only in the $%
\mathbf{x}_{\mathbf{n}}$ -direction, and $\mathbf{u}$ is independent of $%
\mathbf{x}\mathrm{,}$
\begin{equation*}
\mathrm{\delta }x_{\mathbf{n}}^{\prime }=\mathrm{\delta }x_{\mathbf{n}%
}+\left( \frac{\partial u_{\mathbf{n}}}{\partial x_{\mathbf{n}}}\right) _{u}%
\mathrm{\delta }x_{\mathbf{n}}\mathrm{d}s=\mathrm{\delta }x_{\mathbf{n}}.
\end{equation*}%
Deformations due to perpendicular (to $\mathrm{n}$) components of the
velocity cause turns perpendicular to the plane of the initial rectangle,
and thus could affect the area in the second order only.

The acceleration-deformation, $u_{\mathbf{n}}^{\prime }=u_{\mathbf{n}}%
\mathrm{+}a_{\mathbf{n}}\mathrm{d}s,$ acts only in the $\mathbf{u}_{\mathbf{n%
}}$-direction, but can change the area, since%
\begin{equation*}
\mathrm{\delta }u_{\mathbf{n}}^{\prime }=\mathrm{\delta }u_{\mathbf{n}%
}+\left( \frac{\partial a_{\mathbf{n}}}{\partial u_{\mathbf{n}}}\right) _{x}%
\mathrm{\delta }u_{\mathbf{n}}\mathrm{d}s.
\end{equation*}%
Deformations due to perpendicular components of the acceleration have no
effect in the first order.

Thus, the change of the area of the cross-section in each direction along
the trajectory is%
\begin{equation*}
\mathrm{d}S_{\mathbf{n}}=\mathrm{\delta }x_{\mathbf{n}}\mathrm{\delta }u_{%
\mathbf{n}}\left( \frac{\partial a_{\mathbf{n}}}{\partial u_{\mathbf{n}}}%
\right) _{x}\mathrm{d}s.
\end{equation*}%
The change of the volume, $\mathrm{\delta }V,$ which can be represented as a
product of areas of three perpendicular cross-sections, is, obviously,%
\begin{equation}
\mathrm{d\delta }V=\mathrm{\delta }V\mathrm{d}s\cdot \mathrm{div}_{\mathbf{u}%
}\mathbf{a}\mathrm{.}  \label{dVu}
\end{equation}

Note, that

\begin{itemize}
\item due to the choice of the reference frame, the essence of the
derivation is tree-dimensional, i.e., it is valid in the nonrelativistic
case as well;

\item the Lagrangian phase-space and the volume conservation law, (\ref{DVu}%
), can be generalized to multi-particle systems. In this case the volume
will be defined as a product of volumes for each particle, and the rate of
change will be equal to the sum of rates for each particle.

\item the resulting form can be easily transformed to the general reference
frame by replacing the three-dimensional divergence by its four-dimensional
analog. Adding derivatives in the direction of $\mathbf{u}$ cannot change
the value, since $\mathbf{a}$ and $\mathbf{u}$ are always perpendicular to
each other. Thus,%
\begin{equation}
\mathrm{d\delta }V=\mathrm{\delta }V\mathrm{d}s\cdot \mathrm{Div}_{\mathbf{u}%
}\mathbf{a}\mathrm{.}  \label{DVu}
\end{equation}

\item the concept of the four-dimensional divergence is invariant with
respect to changes from flat to curved space, and the effect under
description is just first order. This means, that even if we lost some
generality along the way of derivation, expression (\ref{DVu}) should be
valid in the curved space as well;

\item the divergence of acceleration with respect to velocity is very easy
to find, since the metric tensor is independent of velocity. Indeed,%
\begin{equation}
\mathrm{Div}_{\mathbf{u}}\mathbf{a}\mathrm{=}\frac{\partial a^{\alpha }}{%
\partial u^{\alpha }}=\frac{\partial a_{\alpha }}{\partial u_{\alpha }};
\label{dadu}
\end{equation}

\item the gravitational ($\mathbf{a}\mathrm{=0}$) and electromagnetic ($%
a^{\alpha }=qF^{\alpha \beta }u_{\beta }$) forces always satisfy the
Liouville's theorem, since $F^{\alpha \beta }g_{\beta \alpha }=0$ due to
antisymmetry of $F^{\alpha \beta }$. Also, any force, which is independent
of velocity, satisfies $\mathrm{d\delta }V=0$ automatically.
\end{itemize}

The above list of cases, when $\mathrm{d\delta }V=0$, i.e., the phase-space
volume is conserved, includes all most often encountered cases of dynamical
systems. Besides, it is still incomplete, but also obviously not
all-encompassing. In particular, forces due to the cyclotron damping for a
particle describing the gyrokinetic trajectory cause nonzero contraction of
the phase-volume,
\begin{equation}
\mathrm{Div}_{\mathbf{u}}\mathbf{a}\mathrm{=}\alpha _{r}\left[ F^{\alpha
\beta }F_{\beta \alpha }-3u_{\alpha }F^{\alpha \beta }F_{\beta \gamma
}u^{\gamma }\right] =-\alpha _{r}\left[ H^{2}(2+3w^{2})+E^{2}(1+3w^{2})%
\right] .  \label{divua}
\end{equation}%
As expected, the phase-space volume of the system decreases due to emission
of electromagnetic waves. The rate of the decrease is constant for particles
with non-relativistic perpendicular velocities, $w\rightarrow 0,$ but
increases as $w^{2}$ in the ultra-relativistic limit. The kinetic equation
for this case looks like%
\begin{equation}
\mathrm{d}f+\mathrm{Div}_{\mathbf{u}}\mathbf{a}\mathrm{\cdot }f\mathrm{d}s=0.
\label{dfdv}
\end{equation}

\subsection{The gyrokinetic equation}

At this moment we are ready to construct the gyrokinetic equation with due
account for the cyclotron losses. Starting from the kinetic equation for a
dissipative system, (\ref{dfdv}), we make a transformation to the
gyrokinetic variables, $(x^{\prime \mu },u_{\parallel ,}\widehat{\mu },\phi
) $, in the phase-space. The transformation rules are standard and result in%
\begin{equation}
\frac{\partial f}{\partial x^{\prime \mu }}\mathrm{d}x^{\prime \mu }+\frac{%
\partial f}{\partial u_{\parallel }}\mathrm{d}u_{\parallel }+\frac{\partial f%
}{\partial \hat{\mu}}\mathrm{d}\hat{\mu}+\frac{\partial f}{\partial \phi }%
\mathrm{d}\phi +\mathrm{Div}_{\mathbf{u}}\mathbf{a}\mathrm{\cdot }f\mathrm{d}%
s=0.  \label{gyrokinD}
\end{equation}%
At this point we make an assumption that the motion is almost ideal, i.e.,
the cyclotron emission can be described as a small perturbation, and its
effects can be calculated in zero order in the Larmor radius expansion. As a
result, we can define the perturbed trajectory and the divergence term via
equations (\ref{traject}) and (\ref{divua}). A nice common feature of these
expressions is that they are all independent of the gyrophase ${\phi }$.
This allows to reduce the effective number of variables in the following way.

All coefficients of equation (\ref{gyrokinD}) are independent of ${\phi }$,
while the solution should be periodic in it. If we represent the solution as
a Fourier series, different harmonics of the solution will be completely
independent. Assuming that all ${\phi }$-dependent harmonics ($m>0$) were
either absent initially, or disappeared due to the phase-mixing, it is
possible to consider the behavior of the ${\phi }$-independent part ($m=0$)
alone. It satisfies, obviously,%
\begin{equation*}
\frac{\partial f}{\partial x^{\prime \mu }}\mathrm{d}x^{\prime \mu }+\frac{%
\partial f}{\partial u_{\parallel }}\mathrm{d}u_{\parallel }+\frac{\partial f%
}{\partial \hat{\mu}}\mathrm{d}\hat{\mu}+\mathrm{Div}_{\mathbf{u}}\mathbf{a}%
\mathrm{\cdot }f\mathrm{d}s=0,
\end{equation*}%
which is the gyrokinetic equation with due account for the cyclotron
emission.

The \textquotedblleft magnetic moment\textquotedblright\ is no longer an
exact integral of motion, and is decreasing with time. Note that the rate of
decrease, and the emission-caused reduction of the phase volume produce
terms of the same order and are both equally important. In contrast,
correction to the parallel acceleration seems small and unimportant as
compared to its ideal counterpart. It is certainly small, but in some cases
describes important physical effects. For example, if the parallel momentum
of the particle is conserved, the cyclotron damping of the perpendicular
component of the velocity should be accompanied by the increase of its
parallel velocity, since the relativistic mass of the particle decreases.

\section{The Maxwell equations}

In general, kinetic description of plasmas involves a combination of the
kinetic equation and the Maxwell equations, which describe the evolution of
the collective electromagnetic fields. The same is true for the gyrokinetic
theory. However, since the gyrokinetic equation is written in specific
gyrokinetic variables, and thus yields the distribution function in terms of
these variables, a special procedure of integration in velocity space is
necessary to calculate the source terms of Maxwell equations. This procedure
is presented below.

The general form of the Maxwell's equations in presence of an arbitrary
gravitational field is well known\cite{LL1}. The second subset
\begin{equation}
\frac{1}{\sqrt{-g}}\frac{\partial }{\partial x^{\nu }}\left( \sqrt{-g}F^{\mu
\nu }\right) =\frac{4\pi }{c}j^{\mu },  \label{sp}
\end{equation}%
has a source term,
\begin{equation}
j^{\mu }=c\sum_{\alpha }q_{\alpha }\int u^{\mu }f_{\alpha }\left( \mathbf{x},%
\mathbf{u}\right) \delta V_{u},
\end{equation}%
which is the four-current density, expressed via the distribution function
of particle species, $f_{\alpha }$, the signed particle charge, $q_{\alpha }$
and the element of the volume of the velocity-subspace, $\delta V_{u}.$ Our
current goal is to express $f_{\alpha }\left( \mathbf{x},\mathbf{u}\right) $
in terms of the gyrokinetic distribution function, and $u^{\mu }\delta V_{u}$
- in terms of the gyrokinetic variables.

Starting with $\delta V_{u},$ we note that the four-velocity can be
expressed via Eqs. (\ref{un1})-(\ref{un3}) as
\begin{equation}
u_{\mu }=w\left( l_{\mu }^{\prime }\cos \phi +l_{\mu }^{\prime \prime }\sin
\phi \right) +{u_{\parallel }}l_{\mu }+u_{o}\tau _{\mu }.
\end{equation}%
Taking the box sides along the tetrad vectors, and using Eq.(\ref{orderb}),
we find%
\begin{equation}
\delta V_{u}=\delta \left( \sqrt{u^{\nu }u_{\nu }}-1\right) \mathrm{d}^{4}%
\mathbf{u}=-\frac{w\mathrm{d}w\mathrm{d}\phi \mathrm{d}u_{\parallel }}{u_{o}}%
,  \label{deltavu}
\end{equation}%
where $u_{o}=\sqrt{1+w^{2}+{u_{\parallel }}^{2}},$ and $g^{\prime }=g(%
\mathbf{x}^{\prime })$ is the determinant of the metric tensor in the point $%
\mathbf{x}^{\prime }$, where the tetrad is defined.

That is all we need in the $\lambda =0$ limit, when the differentials in the
expression (\ref{deltavu}) are independent. In the presence of a wave, $%
\lambda \neq 0,$ the gyrokinetic variables are expressed in terms of $\bar{w}%
,\bar{u}_{\parallel }$\ rather than $w\mathrm{,}u_{\parallel }$. However, it
is more convenient to $w\mathrm{,}u_{\parallel }$ as independent variablees,
but transform the distribution function instead. we can make a
transformation of variables, $wu_{\parallel }\phi \rightarrow \bar{w}\bar{u}%
_{\parallel }\phi ,$ and apply its Jacobian. In order $\lambda ^{1}$ we
retain linear corrections only, so that%
\begin{equation}
\frac{w\mathrm{d}w\mathrm{d}\phi \mathrm{d}u_{\parallel }}{u_{o}}=\frac{\bar{%
w}\mathrm{d}\bar{w}\mathrm{d}\phi \mathrm{d}\bar{u}_{\parallel }}{\bar{u}_{o}%
}\left( 1+\frac{\partial \tilde{w}}{\partial \bar{w}}+\frac{\partial \tilde{u%
}_{\parallel }}{\partial \bar{u}_{\parallel }}+\frac{\tilde{w}}{\bar{w}}-%
\frac{\tilde{u}_{o}}{\bar{u}_{o}}\right) .
\end{equation}

As a result, the expression for components of the current density can be
rewritten as
\begin{equation}
j^{\mu }=-\sum_{\alpha }cq_{\alpha }\int \left( w\left( l^{\prime \mu }\cos
\phi +l^{\prime \prime \mu }\sin \phi \right) +{u_{\parallel }}l^{\mu
}+u_{o}\tau ^{\mu }\right) f_{\alpha }\left( \mathbf{x},\mathbf{u}\right)
\frac{w\mathrm{d}w\mathrm{d}\phi \mathrm{d}u_{\parallel }}{u_{o}}.
\label{jm}
\end{equation}%
Further, the distribution function $f_{\alpha }$ is expressed as the
function of the gyrokinetic variables
\begin{equation*}
f_{\alpha }=f_{\alpha }\left( x^{\prime \mu },\hat{\mu},u_{\parallel
}\right) ,
\end{equation*}%
and it is necessary to transform it back to particle coordinates before
integrating, as in Eq.(\ref{jm}) the particle position $\mathbf{x}$, rather
than its gyrocenter position $\mathbf{x}^{\prime },$\ is kept constant while
integrating over the particle velocity. This makes it convenient to rewrite
Equation (\ref{sp}) as
\begin{equation}
\frac{\partial }{\partial x^{\nu }}\left( \sqrt{-g}F^{\mu \nu }\right) =%
\frac{4\pi }{c}j^{\mu }\sqrt{-g}=Q^{\mu }(\mathbf{x}),
\end{equation}%
where the right-hand side is also evaluated at $\mathbf{x}.$ Then
\begin{equation}
Q^{\mu }(\mathbf{x})=4\pi \sum_{\alpha }q_{\alpha }\int \left[ w\left(
l^{\prime \mu }\cos \phi +l^{\prime \prime \mu }\sin \phi \right) +{%
u_{\parallel }}l^{\mu }+u_{o}\tau ^{\mu }\right] f_{\alpha }\left( \mathbf{x}%
-{\sum_{i=1}}\varepsilon ^{i}\mathbf{r}_{i}\right) \sqrt{-g^{\prime }}\frac{w%
\mathrm{d}w\mathrm{d}\phi \mathrm{d}u_{\parallel }}{u_{o}}.
\end{equation}%
Also, it is necessary to keep in mind that the components of the basis
vectors, as well as the relationship between the magnetic moment $\hat{\mu}$
and the orbital velocity $w$, are defined through the electromagnetic field
tensor at the gyrocenter position, and thus should also be transformed back
to the particle position via the inverse transformation
\begin{equation}
x^{\mu }=x^{\prime \mu }+\sum_{i=1}\varepsilon ^{i}r_{i}^{\mu }  \label{it}
\end{equation}%
before integrating.

The gyrokinetic transformation has been found by expansion in orders of $%
\varepsilon$\ and $\lambda$, and this expansion has to be exploited here
again. The short-wavelength wave contributions to the distribution function
can be separated from the slowly changing background by using the Fourier
components as
\begin{equation}
f_{\alpha}\left( x^{\prime\mu},\hat{\mu},u_{\parallel}\right)
=f_{\alpha}^{(0)}+\frac{\lambda}{4\pi^{2}}\int f_{\alpha}^{(1)}\left(
\mathbf{k},\hat{\mu},u_{\parallel}\right) \exp\left[ ik_{\mu} x^{\prime\mu}%
\right] \mathrm{d}^{4}\mathbf{k},
\end{equation}
while $Q^{\mu}(\mathbf{x})$\ should be expanded in powers of $\lambda\ $and $%
\varepsilon$\ as
\begin{equation}
Q^{\mu}(\mathbf{x})=\sum_{i,j=0}\varepsilon^{i}\lambda^{j}Q_{(ij)}^{\mu}.
\end{equation}

The zero-order current and charge density $Q_{(00)}^{\mu }$ is easy to find
\begin{equation}
Q_{(00)}^{\mu }=8\pi ^{2}\sum_{\alpha }\frac{q_{\alpha }^{2}}{m_{\alpha
}c^{2}}\int H^{\dagger }(x^{\mu })\left( \frac{{u_{\parallel }}}{u_{o}}%
l^{\mu }+\tau ^{\mu }\right) f_{\alpha }^{(0)}\left( x^{\mu },\hat{\mu}%
,u_{\parallel }\right) \mathrm{d}\hat{\mu}\mathrm{d}u_{\parallel },
\end{equation}%
where the basis components are taken at the current position. Here and below
we use the following notations:
\begin{equation}
H^{\dagger }(\hat{\mu},x^{\mu })=\frac{w}{q}\frac{\partial w}{\partial \hat{%
\mu}},\hspace{0.1in}H^{\ast }(\hat{\mu},x^{\mu })=\frac{w^{2}}{2q\hat{\mu}},
\end{equation}%
which differ from the magnetic-field invariant due to corrections to the
magnetic moment (found above) in a suitable approximation. The use of $%
H^{\dagger }$ allows to avoid rewriting corresponding corrections in high-$%
\lambda $ contributions to the current density.

\subsection{Order $\protect\lambda^{0}$}

The background distribution function $f_{\alpha}^{(0)}$\ is slowly changing
on the Larmor-radius / gyrotime-scale, and thus can be expanded in the
Taylor series around the particle position
\begin{equation}
f_{\alpha}^{(0)}\left( x^{\prime\mu},\hat{\mu},u_{\parallel}\right)
=f_{\alpha}^{(0)}\left( x^{\mu},\hat{\mu},u_{\parallel}\right) -\varepsilon
r_{1}^{\mu}\frac{\partial f_{\alpha}^{(0)}}{\partial x^{\mu} }%
-\varepsilon^{2}r_{2}^{\mu}\frac{\partial f_{\alpha}^{(0)}}{\partial x^{\mu}
}+\frac{1}{2}\varepsilon^{2}r_{1}^{\mu}r_{1}^{\nu}\frac{\partial^{2}f_{%
\alpha }^{(0)}}{\partial x^{\mu}\partial x^{\nu}}+o(\varepsilon^{2}),
\end{equation}
while the displacements $r_{i}^{\mu}$ are given by the gyrokinetic
transformation as functions of $\left( x^{\prime\mu},\hat{\mu }%
,u_{\parallel}\right) $ and thus should be expanded as well,
\begin{equation}
f_{\alpha}^{(0)}\left( x^{\prime\mu}\right) =f_{\alpha}^{(0)}\left( x^{\mu
}\right) -\varepsilon r_{1}^{\mu}\frac{\partial f_{\alpha}^{(0)}}{\partial
x^{\mu}}+\varepsilon^{2}r_{1}^{\nu}\frac{\partial r_{1}^{\mu}}{\partial
x^{\nu}}\frac{\partial f_{\alpha}^{(0)}}{\partial x^{\mu}}-\varepsilon
^{2}r_{2}^{\mu}\frac{\partial f_{\alpha}^{(0)}}{\partial x^{\mu}}+\frac{1} {2%
}\varepsilon^{2}r_{1}^{\mu}r_{1}^{\nu}\frac{\partial^{2}f_{\alpha}^{(0)} }{%
\partial x^{\mu}\partial x^{\nu}}+o(\varepsilon^{2}).
\end{equation}
Here, finally, all functions in the right-hand side can be regarded as
functions of the particle position $x^{\mu}$. The same expansion should be
applied to the other factor under the integral, namely,
\begin{equation}
G^{\mu}=\left( \sqrt{2qH^{\ast}\hat{\mu}}\left( l^{\prime\mu}\cos
\phi+l^{\prime\prime\mu}\sin\phi\right) +{u_{\parallel}}l^{\mu}+\tau^{\mu
}\right) \frac{H^{\dagger}}{u_{o}}.
\end{equation}

We have
\begin{equation*}
G^{\lambda}\left( x^{\prime\mu},\hat{\mu},u_{\parallel}\right)
=G^{\lambda}\left( x^{\mu},\hat{\mu},u_{\parallel}\right) -\varepsilon
r_{1}^{\mu}\frac{\partial G^{\lambda}}{\partial x^{\mu} }%
+\varepsilon^{2}r_{1}^{\nu}\frac{\partial r_{1}^{\mu}}{\partial x^{\nu} }%
\frac{\partial G^{\lambda}}{\partial x^{\mu}}-\varepsilon^{2}r_{2}^{\mu }%
\frac{\partial G^{\lambda}}{\partial x^{\mu}}+\frac{1}{2}\varepsilon^{2}
r_{1}^{\mu}r_{1}^{\nu}\frac{\partial^{2}G^{\lambda}}{\partial
x^{\mu}\partial x^{\nu}}+o(\varepsilon^{2}),
\end{equation*}
and thus
\begin{multline}
G^{\lambda}f_{\alpha}^{(0)}\left( x^{\prime\mu}\right) =G^{\lambda}
f_{\alpha}^{(0)}\left( x^{\mu}\right) -\left( \varepsilon r_{1}^{\mu
}-\varepsilon^{2}r_{1}^{\nu}\frac{\partial r_{1}^{\mu}}{\partial x^{\nu} }%
+\varepsilon^{2}r_{2}^{\mu}\right) \left( f_{\alpha}^{(0)}\frac{\partial
G^{\lambda}}{\partial x^{\mu}}+G^{\lambda}\frac{\partial f_{\alpha}^{(0)} }{%
\partial x^{\mu}}\right) +  \notag \\
+\varepsilon^{2}r_{1}^{\mu}r_{1}^{\nu}\frac{\partial G^{\lambda}}{\partial
x^{\mu}}\frac{\partial f_{\alpha}^{(0)}}{\partial x^{\nu}}+\frac{1} {2}%
\varepsilon^{2}r_{1}^{\mu}r_{1}^{\nu}\left( f_{\alpha}^{(0)} \frac{%
\partial^{2}G^{\lambda}}{\partial x^{\mu}\partial x^{\nu}}+G^{\lambda }\frac{%
\partial^{2}f_{\alpha}^{(0)}}{\partial x^{\mu}\partial x^{\nu}}\right)
+o(\varepsilon^{2}).
\end{multline}
The first order displacement of the gyrocenter is given by
\begin{equation}
r_{1}^{\mu}=\frac{w}{qH}(l^{\prime\prime\mu}\cos\phi-l^{\prime\mu}\sin
\phi)+\lambda D^{\mu\nu}\tilde{a}_{\nu},
\end{equation}
and, substituting it into the above expression, we find the first-order
current density
\begin{equation}
Q_{(10)}^{\mu}=\sum_{\alpha}\frac{8\pi^{2}q_{\alpha}^{2}}{m_{\alpha}c^{2}}
\int\frac{\hat{\mu}}{\sqrt{H}}\left( l^{\prime\prime\nu}\frac{\partial }{%
\partial x^{\nu}}\left( \frac{{H}^{3/2}f_{\alpha}^{(0)}}{u_{o}}l^{\prime
\mu}\right) -l^{\prime\nu}\frac{\partial}{\partial x^{\nu}}\left( \frac{{H}%
^{3/2}f_{\alpha}^{(0)}}{u_{o}}l^{\prime\prime\mu}\right) \right) \mathrm{d}%
\hat{\mu}\mathrm{d}u_{\parallel}.
\end{equation}
Similarly, it is necessary to solve the equation (\ref{forr2}) for $r_{2}
^{\mu}$ in order to calculate the second-order current density. The form of
this equation (for $\lambda=0$) is such that $r_{2}^{\nu}$ may have
components proportional to trigonometric functions of $\phi$ and of $2\phi.$
However, due to the integration in $\phi,$ only the first type of terms will
produce a non-zero contribution to the second-order current density. Thus,
it is sufficient to solve for $r_{21}^{\nu},$ which satisfies
\begin{equation}
qr_{21}^{\nu}F{_{\mu\nu}-}\left( {u_{\parallel}}l_{\nu}+u_{o}\tau_{\nu
}\right) \frac{\partial r_{1}^{\nu}}{\partial x^{\prime{\mu}}}=0,
\end{equation}
or
\begin{equation}
r_{21}^{\lambda}=\frac{1}{q}D^{\lambda\mu}\bar{u}_{\nu}\frac{\partial
r_{1}^{\nu}}{\partial x^{\prime{\mu}}}=\frac{1}{q}D^{\lambda\mu}\left( {%
u_{\parallel}}l_{\nu}+u_{o}\tau_{\nu}\right) \frac{\partial}{\partial
x^{\prime{\mu}}}\frac{w}{qH}(l^{\prime\prime\nu}\cos\phi-l^{\prime\nu}\sin
\phi).
\end{equation}
Using this expression we get \ \
\begin{multline}
Q_{(20)}^{\mu}=-4\pi^{2}\sum_{\alpha}m_{\alpha}c^{2}\int\hat{\mu }\mathrm{d}%
\hat{\mu}\mathrm{d}u_{\parallel}\left\{ \frac{\partial }{\partial x^{\nu}}%
\left[ \frac{\left( l^{\prime\nu}l^{\prime\lambda
}+l^{\prime\prime\nu}l^{\prime\prime\lambda}\right) }{H}\frac{\partial }{%
\partial x^{\lambda}}\left( Hf_{\alpha}^{(0)}\frac{\bar{u}^{\mu}}{u_{o} }%
\right) \right] -\right. \\
\left. -2D^{\lambda\varsigma}\bar{u}_{\nu}\left[ \frac{\partial}{\partial
x^{\varsigma}}\left( \frac{l^{\prime\prime\nu}}{\sqrt{H}}\right) \frac{%
\partial}{\partial x^{\lambda}}\left( \frac{H^{3/2}f_{\alpha}^{(0)} }{u_{o}}%
l^{\prime\mu}\right) -\frac{\partial}{\partial x^{\varsigma}}\left( \frac{%
l^{\prime\nu}}{\sqrt{H}}\right) \frac{\partial}{\partial x^{\lambda} }\left(
\frac{H^{3/2}f_{\alpha}^{(0)}}{u_{o}}l^{\prime\prime\mu}\right) \right]
\right\} .
\end{multline}

\subsection{Order $\protect\lambda^{1}$}

First we calculate the transform to the particle position variables
\begin{equation}
f_{\alpha }^{(1)}\left( \mathbf{x},\hat{\mu},u_{\parallel },\phi \right) =%
\frac{1}{4\pi ^{2}}\int f_{\alpha }^{(1)}\left( \mathbf{k},\hat{\mu}%
,u_{\parallel }\right) e^{i\mathbf{kx}}\exp \left[ -ik_{\mu }{\sum_{s}}%
\varepsilon ^{s}r_{s}^{\mu }\right] \mathrm{d}^{4}k.
\end{equation}%
Retaining in the exponent terms of order $\varepsilon ^{0}$ only [while $\
k\varepsilon \sim O(1)$], we get
\begin{equation}
f_{\alpha }^{(1)}\left( \mathbf{x}\right) =\frac{1}{4\pi ^{2}}\int f_{\alpha
}^{(1)}\left( \mathbf{k}\right) e^{i\mathbf{kx}}\exp \left[ -i\varepsilon
k_{\mu }\left( \frac{w}{qH}(l^{\prime \prime \mu }\cos \phi -l^{\prime \mu
}\sin \phi )+\lambda D^{\mu \nu }\tilde{a}_{\nu }\right) \right] \mathrm{d}%
^{4}k.  \label{fa1}
\end{equation}%
The last term in the exponent can be neglected in this order, and the rest
can be expanded into the Fourier series using Eq.(\ref{tojd}) as
\begin{equation}
f_{\alpha }^{(1)}\left( \mathbf{x}\right) =\frac{1}{4\pi ^{2}}\int f_{\alpha
}^{(1)}\left( \mathbf{k}\right) e^{i\mathbf{kx}}\sum_{n=-\infty }^{+\infty
}J_{n}\left( \xi \right) e^{in(\phi -\phi _{0})}\mathrm{d}^{4}k,
\end{equation}%
where $\xi =k_{\bot }\rho ,\ \rho =\varepsilon w/qH=\varepsilon \sqrt{2\hat{%
\mu}/qH},\ \ k_{\bot }=\sqrt{\left( k_{\nu }l^{\prime \nu }\right)
^{2}+\left( k_{\nu }l^{\prime \prime \nu }\right) ^{2}},$ as before,$\ \tan
\phi _{0}=\left( k_{\mu }l^{\prime \prime \mu }\right) /\left( k_{\mu
}l^{\prime \mu }\right) $ . Then,
\begin{equation}
Q_{(01)}^{\mu }=-2\sum_{\alpha }\frac{q_{\alpha }^{2}}{m_{\alpha }c^{2}}\int
H^{\dagger }\frac{\mathrm{d}\hat{\mu}\mathrm{d}u_{\parallel }}{u_{o}}\int
\mathrm{d}^{4}kf_{\alpha }^{(1)}\left( \mathbf{k}\right) e^{i\mathbf{kx}%
}\left\{ \left( {u_{\parallel }}l^{\mu }+u_{o}\tau ^{\mu }\right)
J_{0}\left( \xi \right) +2i\varepsilon \hat{\mu}b^{\mu \nu }k_{\nu }\frac{%
J_{1}\left( \xi \right) }{\xi }\right\} .
\end{equation}%
Here we have again used $H^{\dagger }$ instead of $H$ to take into account
the $\hat{\mu}$-related corrections in future orders.

\section{Conclusion}

This work has been conducted via the cooperation program between
the Trieste University, Italy, and the Budker Institute of Nuclear
Physics, Novosibirsk, Russia, and supported by a grant of INdAM
(Istituto Nazionale di Alta Matematica, Italy).

\end{document}